\documentclass[12pt]{article}
\usepackage{xcolor,pstool,enumitem,tikz,tensor,amsthm,mathtools,soul,braket,amsbsy,amsmath,amsfonts,amssymb,cite,float,esint}

\usepackage[most]{tcolorbox}
\tcbset{grayeq/.style={colback=gray!15, colframe=gray!15, boxrule=0pt, sharp corners,
  boxsep=0pt, left=4pt, right=4pt, top=3pt, bottom=3pt, grow to left by=4pt, grow to right by=4pt,
  before skip=\abovedisplayskip,
  after={\par\nointerlineskip\addvspace{\belowdisplayskip}\noindent\ignorespacesafterend}}}
\newtcolorbox{grayequation}{ams equation, grayeq}
\newtcolorbox{grayalign}{ams align, grayeq}

\usepackage{bbold}
\usepackage{verbatim}
\DeclareMathAlphabet{\mathbbold}{U}{bbold}{m}{n}
\DeclareMathAlphabet{\mathbb}{U}{msb}{m}{n}

\usetikzlibrary{decorations.pathmorphing,decorations.markings,arrows.meta}
\tikzset{
  flow/.style={decoration={markings, mark=at position #1 with {\arrow[scale=1.0]{Stealth}}}, postaction={decorate}},
  flow/.default=0.55,
  bcut/.style={decorate, decoration={snake, amplitude=0.7pt, segment length=6pt, pre length=2pt, post length=2pt}},
  cdraw/.style={line width=0.8pt, baseline=-2.4pt},
}

\usepackage[unicode,colorlinks = true,
            linkcolor = blue!70!black,
            urlcolor  = red!70!black,
            citecolor = green!55!black,
            anchorcolor = blue!70!black]{hyperref}
\usepackage[margin=0.8in]{geometry}
\usepackage{placeins}

\setlist[enumerate]{listparindent=\parindent,parsep=0pt}

\renewcommand*{\Re}{\textsf{Re}}
\renewcommand*{\Im}{\textsf{Im}}
\def\di{\mathrm{d}}

\begin{document}
%%%%%%%%%%%%%%%%%%%%%%%%%%%%%%%%%%%%%%%%%%%%%%%%
\begin{titlepage}
\hfill \\
\vspace*{15mm}
\begin{center}
{\LARGE \bf Tunneling from an oscillating state \\ in quantum mechanics}
\vspace*{15mm}

{\large Oliver Janssen$^{1}$, Matthew Kleban$^{2}$ and Cameron Norton$^{2}$}

\vspace*{8mm}

{\small
$^{1}$Laboratory for Theoretical Fundamental Physics, EPFL, 1015 Lausanne, Switzerland \\
\vspace*{2mm}
$^{2}$Center for Cosmology and Particle Physics, New York University, New York, NY 10003, USA
}

\vspace*{0.7cm}
\end{center}
\begin{abstract}

\noindent \normalsize We study the decay of general initial states out of a metastable potential well in quantum mechanics. We provide a closed-form expression for the probability current that tunnels through the barrier in terms of the resonant states into which the initial state can be decomposed. All ingredients are computed analytically to first subleading order in the semiclassical limit. Specializing to a coherently-oscillating state, we derive an equivalent semiclassical approximation to the current via the path integral, which circumvents the decomposition into resonant states. We test the accuracy of the approximation by comparing it to a numerical solution of the Schr\"odinger equation.
\end{abstract}

\vspace{1cm}

\today

\end{titlepage}
%%%%%%%%%%%%%%%%%%%%%%%%%%%%%%%%%%%%%%%%%%%%%%%%

\tableofcontents

%%%%%%%%%%%%%%%%%%%%%%%%%%%%%%%%%%%%%%%%%%%%%%%%
\section{Introduction}
%%%%%%%%%%%%%%%%%%%%%%%%%%%%%%%%%%%%%%%%%%%%%%%%
Systems in metastable states can decay via quantum tunneling across classically forbidden regions. When the initial state is the perturbative vacuum of the metastable region, this is a standard calculation using the WKB approximation or semiclassical path integral methods \cite{Coleman:1977py, Callan:1977pt}. However, in many physically relevant situations, the system is not initially in the metastable ground state. In superconducting circuits, a current-biased Josephson junction can tunnel out of a zero-voltage metastable state via macroscopic quantum tunneling. Experimentally, it has been verified that the decay rate depends sensitively on whether the system begins in the ground state or an excited state of the tilted washboard potential \cite{PhysRevLett.55.1543, PhysRevB.35.4682}.

In cosmology, related situations arise in multi-field inflation models where one field evolves dynamically while another direction in field space supports tunneling. In such scenarios, tunneling can occur from the slow-roll trajectory to a minimum in a different region of the potential, so the initial state has non-vanishing momentum and is not localized near a minimum \cite{Garriga:2016bvm}. In \cite{Kleban:2025pbh}, a simple approximation was developed to describe this tunneling rate as a function of time; a more recent method is \cite{Yamada:2026xsd}. A second cosmological example is an oscillating  scalar field, such as axion dark matter. If the local minimum about which the field is oscillating is not the global minimum of the potential, the field can decay by bubble nucleation.  One expects that coherent oscillations around this minimum exponentially enhance the decay rate.  Despite the fundamental nature of this problem, a general framework for describing tunneling out of non-vacuum states is still incomplete. In quantum field theory, different approaches to this problem yield results that disagree even at the level of the exponent \cite{Krais-Vakkuri_1996, Darme:2019ubo}.

When the initial state is a superposition of energy eigenstates, the different energy components tunnel at different rates, and their interference produces oscillations in the probability current that are absent for a single component. Although progress has been made on understanding this \cite{PhysRevD.111.096009}, we are unaware of a simple equation which allows one to straightforwardly approximate the tunneling rate, including its time dependence on sub-exponential timescales. We were inspired by the work of \cite{Lin:2025bjn,Lin:2025wgc}, which uses resonant states to obtain the shape of the tunneling probability as a function of time. In their discussion, several coefficients were fit to the numerically time-evolved wave function. By contrast, our result \eqref{formula} is a closed-form expression for the probability current through the barrier that captures both the exponential decay of individual resonances and the interference between different resonant states.

The outline of this paper is as follows. We first express the outgoing probability current for a general initial state localized in the metastable well in terms of resonant states (\S\ref{resstatesec}-\ref{probcurrentsec}). Second, we recall how the ingredients of this formula are computed semiclassically (\S\ref{WKBsec} and Appendices \ref{excitedWKB}-\ref{lowlyingWKB}, where we also introduce our notation). Third, specializing to a specific family of coherent initial states, we derive two equivalent approximations to the associated current: one decomposes the initial state into resonant states as in \S\ref{resstatesec} and uses a saddle-point approximation on the sum (\S\ref{decompsec}), while the other uses a semiclassical approximation to the time-evolved state expressed as a path integral (\S\ref{PIsec}). This general result is expressed in terms of an implicit solution to a saddle-point equation, which can be solved (perhaps, numerically) for a given potential. Perturbation theory near the time of maximal tunneling (\S\ref{nearmaxsec}) shows that tunneling occurs in narrow bursts near the classical turning point. In \S\ref{smalloscsection} we study \textit{small} oscillations, for which the saddle point only explores the quadratic regime of the potential around the metastable minimum. There the saddle-point equation can be solved explicitly and a simple closed-form expression for the time-dependent tunneling current is obtained. In \S\ref{generalcoherentsec} we give the result for another family of coherent initial states. Finally, we verify our results numerically in an example (\S\ref{sec:Numerics}). Our main analytic results, \eqref{formula}-\eqref{Jexpressiongeneral}-\eqref{joftfinalGarg}, are highlighted.

In this work we focus on one-dimensional quantum mechanics, leaving the extension to multiple dimensions and quantum field theory for the future. Our path integral calculations follow the method described in \cite{Janssen:2026ybl}.

%%%%%%%%%%%%%%%%%%%%%%%%%%%%%%%%%%%%%%%%%%%%%%%%
\section{Tunneling from general trapped states}
\begin{figure}[ht]
    \centering
    \includegraphics[width=0.63\linewidth]{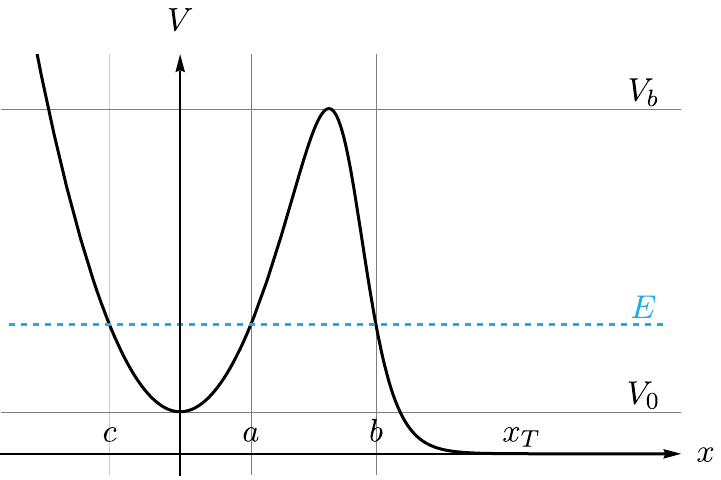}
    \caption{A generic potential $V$ assumed in our analysis. A well (left) is shown separated from an asymptotically free region (right) by a potential barrier. The dashed line shows a representative energy $E$. The classical turning points for this energy are $c$ and $a$ on the left and right in the well, and $b$ on the right of the barrier. The start of the true vacuum, or free region, is denoted by $x_T$.}
    \label{fig:potential}
\end{figure}

%%%%%%%%%%%%%%%%%%%%%%%%%%%%%%%%%%%%%%%%%%%%%%%%
Consider a potential with a false vacuum (a well) separated from a true vacuum (an asymptotically free region) by a barrier, shown in Fig.~\ref{fig:potential}. 
An initial state $\psi(x)$ which is localized in the well with energy well below the top of the barrier is classically stable, but quantum mechanically will slowly decay into the true vacuum. To explicitly calculate its tunneling behavior, one could expand $\psi(x, t)$ in a basis of the exact energy eigenstates of the Hamiltonian of the system, $H$, and evolve each in time. If we do not have access to these exact energy eigenstates, another approach is  to find a set of approximate energy eigenstates which accurately describe the physics of the problem.

%%%%%%%%%%%%%%%%%%%%%%%%%%%%%%%%%%%%%%%%%%%%%%%%
\subsection{Resonant state expansion} \label{resstatesec}
%%%%%%%%%%%%%%%%%%%%%%%%%%%%%%%%%%%%%%%%%%%%%%%%
In the limit where the barrier is taken to be infinitely high, the energy eigenstates are bound states localized in the well, with quantized energies and no tunneling. When the barrier is finite and the true vacuum extends to infinity -- the limit which corresponds to physically relevant situations -- the spectrum is continuous, and generic energy eigenstates are delta-function normalizable scattering states with significant relative support in the asymptotic region. However, near each bound-state energy of the confined problem, the full Hamiltonian develops a resonance, which becomes exponentially narrow in the semiclassical limit. States with energies in the vicinity of these resonances have large relative amplitude in the well and only small outgoing flux into the true vacuum. This sparse set of states which are localized in the well contains the relevant eigenstates for the decomposition of an initial state localized in the well. We review this in Appendices \ref{excitedWKB}-\ref{lowlyingWKB}.

When numerically solving the Schr\"odinger equation, however, one must cut off the true vacuum at some finite distance from the barrier. Doing this quantizes the energy, and leaves only a discrete set of normalizable energy eigenstates. At most a small subset of these states corresponds to the quasi-bound states localized in the well. Since the associated resonances are exponentially narrow in the semiclassical limit, the corresponding resonant energies occupy very small intervals in the continuum of states. Unless the cutoff is taken very far away, the discretized spectrum fails to resolve these narrow resonances, and the quasi-bound states with considerable amplitude in the well are missed.

Instead, one can consider resonant (Gamow-Siegert) states, defined by imposing outgoing boundary conditions to the right of the barrier \cite{Gamow1928, Siegert1939}; see also \cite[\S 134]{Landau:1991wop}. Physically, this mimics the situation in which the asymptotically free region extends to infinity, since the wave function is not allowed to reflect from a distant boundary and tunnel back into the well. As a result, probability can continuously flow out of the metastable region.

Allowing net probability flux through the boundary implies that the
corresponding Hamiltonian is non-Hermitian. The continuum of scattering states is replaced with a discrete spectrum of resonant states with complex energies localized in the well. These states satisfy
\begin{equation} \label{resonantenergies}
    \widetilde H \psi_n(x) = \varepsilon_n \psi_n(x) \,, \quad \varepsilon_n = E_n - i \frac{\hbar \Gamma_n}{2} \,,
\end{equation}
where $\widetilde H$ is the non-Hermitian Hamiltonian and $\Gamma_n > 0$.\footnote{Notice that as differential operators, $\widetilde H = H$. However, these operators act on different spaces -- decaying functions on the left and purely outgoing on the right, vs.~Schwartz functions on the real line. The tilde denotes this difference. In numerical studies, one effectively imposes the outgoing boundary condition by adding a complex absorbing term to $H$ at a finite point deep in the free region, as we explain in \S\ref{sec:Numerics}.} As the barrier height is increased, the real parts of the energies $E_n$ approach the bound-state energies of the confined problem, and the widths $\Gamma_n$ tend to zero. The $\psi_n$ that can be uniformly rotated to have small imaginary part are those that are well-localized in the well and are slowly-decaying, with a lifetime of order $1/\Gamma_n$.

We propose to expand the initial state in the ``basis" of resonant states.
Because $\widetilde H$ is not Hermitian, the $\psi_n$ do not automatically possess several properties that make the eigenstates of a Hermitian operator a convenient basis: normalizability, completeness, and orthogonality. We discuss each of these in turn and argue that, despite these subtleties, the expansion remains well-justified for the initial state we consider. 

The eigenfunctions $\psi_n$ are not normalizable in the usual sense because the outgoing boundary condition leads to exponential growth deep in the free region. However, they can be assigned a physically meaningful normalization in our case, since we are interested only in states localized in the well where the resonant wave functions remain finite and closely approximate the quasi-bound states of the original Hermitian Hamiltonian $H$.
We therefore fix their overall normalization by requiring
\begin{equation} \label{psi_norm}
    \int_{-\infty}^{x_T} \di x \, |\psi_n(x)|^2 = 1 \,,
\end{equation}
where $x_T$ lies to the right of the barrier where the free region starts.

There is also no guarantee that the $\psi_n$ form a complete basis of the Hilbert space. In general, a complete spectral decomposition requires both resonant contributions and a non-resonant continuum component. This is formalized in the Berggren completeness relation, which augments the discrete set of resonant states with an integral over a continuum contour in the complex energy plane  \cite{Berggren1968, Garcia-Calderon:1976omn}. However, for states localized in the potential well, the overlap with the non-resonant continuum is small.  In these cases, decomposing the initial state as a sum of resonant states is a useful approximation. This approximation becomes less accurate for states with significant support at higher energies, where contributions from the non-resonant continuum are no longer negligible (and at very early and very late times, as we discuss later). With these caveats we can thus approximate the initial wave function as 
\begin{equation} \label{resexp}
    \psi(x) = \sum_n c_n \psi_n(x) \,.
\end{equation}

Because the $\psi_n$ are not exactly orthogonal, the coefficients $c_n$ cannot be obtained by simple projection. Taking the inner product of the expansion
with $\psi_m$ gives
\begin{equation}
\langle \psi_m | \psi \rangle =
\sum_n c_n \langle \psi_m | \psi_n \rangle \,,
\end{equation} 
where the inner product is defined by integration over the region $(-\infty, x_T]$.  Since $ \langle \psi_m | \psi_n \rangle \neq 0$ in general for $m \neq n$, the sum does not collapse and we instead define the overlap matrix
\begin{equation}
    O_{mn} \equiv \langle \psi_m | \psi_n \rangle \,.
\end{equation}
The coefficients are obtained by inverting this matrix,
\begin{equation}
c_n = \sum_m (O^{-1})_{nm}\,\langle \psi_m | \psi \rangle \,.
\end{equation}
In practice, the overlap matrix is very nearly diagonal in the narrow-resonance regime. Taking $x_T$ as the right boundary and using the WKB expressions reviewed in the Appendices, one finds
\begin{equation}
|\langle \psi_m |\psi_n \rangle| \sim e^{-(S_n + S_m)/\hbar} \qquad  (n \neq m),
\end{equation}
where $S_k$ is the WKB action for tunneling at energy $E_k$.

Once we have decomposed $\psi$ into resonant states, its time evolution under the complex Hamiltonian is straightforward. The $\psi_n$ evolve under $\widetilde H$ as 
\begin{equation} \label{eigTimeEvolve}
    \psi_n(x, t) = \psi_n(x) \, e^{-i \varepsilon_n t/\hbar} \,,
\end{equation}
so the time evolution of $\psi$ is
\begin{equation} \label{psiHtildeevolution}
    \psi(x, t) = \sum_n c_n \psi_n(x) \, e^{-i E_n t/\hbar} e^{-\Gamma_n t /2} \,.
\end{equation}
 Of course, what we ultimately want is the time evolution of $\psi$ generated by the original Hamiltonian $H$ (for which the time evolution is unitary and time-symmetric when the initial state is real, in contrast to \eqref{psiHtildeevolution}). In \S\ref{sec:Numerics} we show that evolving the resonant states with $\widetilde H$ provides an excellent approximation to the evolution under $H$ in an appropriate intermediate time regime. In the semiclassical limit there is an alternative way of thinking about the resonant states, which we present in Appendix~\ref{resonanttimeevolutionsec}.

%%%%%%%%%%%%%%%%%%%%%%%%%%%%%%%%%%%%%%%%%%%%%%%%
\subsection{Probability current} \label{probcurrentsec}
%%%%%%%%%%%%%%%%%%%%%%%%%%%%%%%%%%%%%%%%%%%%%%%%
We now turn to the time dependence of the tunneling process. This can be obtained by computing the probability current associated with $\psi(x,t)$, 
\begin{equation} \label{currentgeneralexpr}
    j(x,t)=\frac{\hbar}{m}\,\Im \left( \psi(x,t)^* \, \partial_x\psi(x,t) \right) \,.
\end{equation}
The rate at which probability leaves the well is given by the outgoing flux through the point $x_T$, which lies to the right of the barrier where the free region begins. For $x > x_T$, the resonant states take the form 
\begin{equation}\label{psiOutgoing}
    \psi_n(x) = A_n \, e^{ik_n (x-x_T)/\hbar} \,, \quad k_n = \sqrt{2m \left( \varepsilon_n - V(x_T) \right)} \,, \quad A_n \equiv |A_n| \, e^{i \zeta_n/\hbar} \,.
\end{equation}
The momentum $k_n$ is complex because $\varepsilon_n$ is complex. From \eqref{resonantenergies} we see that resonant states grow exponentially at large $x$. The probability current carried by the $n$th
resonant state at $x_T$ is
\begin{equation}
    j_n(x_T,t) = \frac{|A_n|^2}{m} \, \Re \, k_n \, e^{-\Gamma_n t} \,.
\end{equation}
We can find $|A_n|$ in terms of $\varepsilon_n$ by requiring that this outgoing flux equals the rate of decrease of probability in the well. We use the continuity equation to relate the flux to the decay width
$\Gamma_n$,
\begin{equation}
    \partial_t|\psi_n|^2+\partial_x j_n = 0 \,.
\end{equation}
Define the probability to remain in the false vacuum region by
\begin{equation}
P_n(t) \equiv \int_{-\infty}^{x_T} \di x \, |\psi_n(x,t)|^2 = e^{-\Gamma_n t} \,,
\end{equation}
where we have used \eqref{eigTimeEvolve} and $P_n(0) = 1$ by \eqref{psi_norm}. Integrating the continuity equation over $(-\infty,x_T]$, on the other hand, gives
\begin{equation}
\dot P_n(t) = - j_n(x_T,t) \,.
\end{equation}
Therefore
\begin{equation}
   |A_n| = \sqrt{\frac{m \Gamma_n}{\Re \, k_n}} \,.
\end{equation}
The phase $\zeta_n$ of $A_n$ is not determined by this argument, and, as far as we can tell, must be determined independently in general.

Now we can compute the probability current for the state $\psi$ at $x_T$, by writing $\psi$ as a sum of eigenstates of $\widetilde H$ of the form in Eq.~\eqref{psiOutgoing}. If we write $c_n = |c_n| e^{i\vartheta_n}$ and $\delta_n\equiv\arg k_n$, then
\begin{grayalign}
    &j(x_T,t) = \sum_n |c_n|^2 \, \Gamma_n \, e^{-\Gamma_n t}
    \label{formula} \\
    &+\sum_{n\neq n'} |c_n c_{n'} k_{n'}|\!\left({\frac{\Gamma_n \Gamma_{n'}}{\Re \,k_n \, \Re \, k_{n'}}} \right)^{1/2}\!
    \, \cos\!\Big(\frac{E_{n}-E_{n'}}{\hbar}t+(\vartheta_{n'}-\vartheta_n) + \frac{\zeta_{n'} - \zeta_n}{\hbar} + \! \delta_{n'} \!
    \Big) e^{-(\Gamma_{n} \!+\Gamma_{n'}\!)t/2} \,. \notag
\end{grayalign}
The first line represents the time-averaged current, while the second contains the oscillations arising from interference between the resonant states. Interference can have dramatic effects on the current, as we will see.

All we have assumed so far is that the resonant states in \eqref{resonantenergies} have been found, into which the initial state decomposes as in \eqref{resexp}, and we have written the current in terms of $\{ (E_n,\Gamma_n, \zeta_n) \}$. Of course, the usefulness of this formula rests upon being able to decompose an initial state in terms of resonant states. This description becomes incomplete at high energies or very early times. For a real initial state the true current $j(x,t) \sim t$ as $t \to 0$.  In particular, it is identically zero at $t = 0$, while the resonant expansion \eqref{formula} is finite at $t = 0$. The correct early time dynamics arises from cancellations between resonant and non-resonant (continuum) components of the full spectral decomposition \cite{FondaGhirardiRimini1978}. Accordingly, Eq.~(\ref{formula}) should be understood as providing an accurate description of the current only after a brief initial transient period (of order one oscillation at a representative energy in the sum). It is also the case that at very late times $t \gtrsim \log \beta^4/\Gamma_0$, the decay is no longer exponential \cite{Khalfin,Raczynska:2018ofh}.\footnote{Here $\beta = (E_0 - E_{\textsf{min}})/(\hbar \Gamma_0)$ is assumed to be large, with $E_{\textsf{min}}$ the energy below the metastable ground state energy $E_0$ at which the energy density $|\langle E | \psi_{t = 0} \rangle|^2$ starts deviating significantly from the Breit-Wigner profile. If we estimate this like in Appendix~\ref{excitedWKB} as $\mathcal{O}(\hbar/T(E_0))$, then $\log \beta^4 \sim 8 S_0/\hbar$.}
We will now derive approximations for the resonance energies $E_n$, widths $\Gamma_n$ and phases $\zeta_n$ in the semiclassical limit.

%%%%%%%%%%%%%%%%%%%%%%%%%%%%%%%%%%%%%%%%%%%%%%%%
\subsection{Semiclassical limit} \label{WKBsec}
%%%%%%%%%%%%%%%%%%%%%%%%%%%%%%%%%%%%%%%%%%%%%%%%
We can derive an expression for the complex energies and phases in the semiclassical limit by imposing purely outgoing boundary conditions on the WKB wave functions, reviewed in Appendices \ref{excitedWKB}-\ref{lowlyingWKB}, where our notation is also introduced. In the free region $x \geq b + \Delta_b$ to the right of the barrier, the energy eigenstates take the form in Eq.~\eqref{psifreegeneral}. To impose outgoing boundary conditions, we convert the cosines into exponentials, and declare that the coefficient of the $e^{-i\int k/\hbar }$ term vanishes. This sets
\begin{equation} \label{WKBoutgoingBC}
    2A\theta^2 = i \,.
\end{equation}
For (complex) energies close to where the WKB quantization conditions are met, $\varphi(E_n) = \pi(n+1/2)$, we can expand $A$ as in \eqref{Aclosetoresonance}. Solving for $E = \varepsilon_n$ gives
\begin{equation} \label{complexWKBenergy}
    \varepsilon_n = E_n - i \frac{\hbar \Gamma_n}{2} \,, \quad  \Gamma_n = \frac{1}{g_n T_n} e^{-2 S_n/\hbar} \,.
\end{equation}
The phase $\zeta_n = \hbar \, \arg A_n$ is read off from \eqref{psiXToutsideGeneral},
\begin{equation} \label{AnphaseSC}
    \zeta_n = \int_{b_n}^{x_T} k_n + \frac{\pi \hbar}{4} \,,
\end{equation}
where here and for the remainder of the main text, we have multiplied all resonant states $\psi_n(x)$ by $(-1)^n$ compared to our convention in the Appendices (see e.g.~\eqref{resonantfull}). These results are then used to approximate the probability current in Eq.~\eqref{formula}. We point out that the semiclassical results \eqref{complexWKBenergy}-\eqref{AnphaseSC} -- for $n \gg 1$ where $g_n \to 1$, and for $n = 0$ where $g_0 = \sqrt{e/\pi}$ -- can also be derived by a path integral approach \cite{Janssen:2026ybl}.

At this stage, we can give a rather general expression for the average probability that leaks out of the well per unit time:
\begin{equation}
    \overline{\Delta P} \equiv \frac{1}{T} \int_0^T \di t ~ j(x_T,t) \approx \overline{\Gamma} \,, \quad \overline{\Gamma} = \sum_n |c_n|^2 \Gamma_n \,. \label{Gammabardef}
\end{equation}
This approximation is valid for times much after the initial transient period and much before the state has decayed significantly (that is, a time of order $1/\overline{\Gamma})$. In the semiclassical limit, this time window is exponentially large compared to the classical oscillation time for typical levels in the sum \eqref{formula}.

If we further assume that for the relevant terms in  the sums \eqref{formula}, the energies are well-approximated by their  expressions for a quadratic well $E_n = \hbar \omega(n+1/2)$ (where $\omega/2\pi$ is the frequency of oscillations around the minimum) we can approximate
\begin{equation} \label{DeltaP1}
    \Delta P = \frac{2 \pi}{\omega} \overline{\Gamma}
\end{equation}
for the probability that leaks out during a single oscillation time $2\pi/\omega$.
%In what follows, for simplicity, we will assume the quadratic approximation to the energies $E_n$ is valid. We comment on going beyond this approximation at the end of \S\ref{nearmaxsec}.

%%%%%%%%%%%%%%%%%%%%%%%%%%%%%%%%%%%%%%%%%%%%%%%%
\section{Tunneling from coherent states} \label{coherentsec}
%%%%%%%%%%%%%%%%%%%%%%%%%%%%%%%%%%%%%%%%%%%%%%%%
We now specialize to coherent initial states and seek an approximation to both the total probability that leaks out during one oscillation period and the time window over which this leakage effectively occurs. 

A particular family of coherent states, whose decay we will focus on in \S\ref{decompsec}-\S\ref{smalloscsection}, can be motivated by expanding the potential to quadratic order near the center of the metastable well at $x \equiv 0$ where $V \equiv 0$:
\begin{equation} \label{potentialcorrections}
    V(x) = \frac{m \omega^2}{2} x^2 + \mathcal{O}\left( \lambda x^3 \right) \quad \text{as } x \to 0 \,.
\end{equation}
In our decomposition \eqref{resexp} of the initial state into resonant states of the well, we make the choice
\begin{equation} \label{cncoherent}
    c_n = e^{-\alpha^2/2} \frac{\alpha^n}{\sqrt{n!}} \,, \quad \alpha \geq 0 \,.
\end{equation}
For low-lying energies $E_n \approx \hbar \omega(n+1/2)$ and in the quadratic part of the metastable region, the resonant states take the approximate form of the harmonic oscillator energy eigenstates
\begin{equation}
	\psi_n(x) = \mathcal{N}_n \, e^{-(x/d)^2/2} H_n(x/d) \,, \quad \mathcal{N}_n = \frac{1}{\sqrt{d \sqrt{\pi} \, 2^n n!}} \,, \quad d^2 = \frac{\hbar}{m \omega} \,.
\end{equation}
The initial state is then approximately
\begin{equation} \label{initialstateaalpha}
    \psi(x,0) = \sum_{n=0}^\infty c_n \psi_n(x) = \frac{1}{( \pi d^2 )^{1/4}} \exp \left( - \frac{(x-a)^2}{2d^2} \right) \,, \quad a = \sqrt{2}d \alpha \,,
\end{equation}
provided $a \ll \ell_{\textsf{NL}}$, where $\ell_{\textsf{NL}} \sim m \omega^2/\lambda$ denotes the scale at which corrections to the quadratic approximation \eqref{potentialcorrections} become important. At early times $t \ll t_\textsf{dec}$ when we can neglect the decohering effect of higher-order terms in the potential\footnote{\label{decoherenceftnote}A cubic correction as in \eqref{potentialcorrections} would result in a decoherence time of order $t_\textsf{dec} \sim m^2 \omega^3/\lambda^2 a^2$. One can estimate this time by requiring the additional phase build-up $\partial_n^2 E_n (\Delta n)^2 t/\hbar$ to become $\mathcal{O}(1)$. Here $\Delta n \sim \alpha$ is the width of \eqref{cncoherent} in $n$ at large $\alpha$ (see also footnote \ref{Gaussianfootnote}), and $\partial_n^2 E_n \sim \lambda^2 \hbar^2/m^3\omega^4$ by second order perturbation theory. The classical timescale $t_\textsf{dec}$ is itself much smaller than the quantum timescale $1/\bar{\Gamma}$ at which a significant part of the probability in the well has decayed.}, the time-evolved state in the metastable region is approximately
\begin{equation}
    \psi(x,t) = \frac{1}{( \pi d^2 )^{1/4}} \exp \left( - \frac{(x-a \cos \omega t)^2}{2d^2} \right) e^{i \phi(x,t)} \,, \quad \phi(x,t) = \frac{a \sin \omega t}{2d^2} \left( a \cos \omega t - 2x \right) - \frac{\omega t}{2} \,.
\end{equation}
Therefore, for $a \ll \ell_\textsf{NL}$ and $t \ll t_\textsf{dec}$, the initial state with resonant-state weights \eqref{cncoherent} retains its coherence over time, motivating the choice \eqref{cncoherent}.

One can nevertheless study states with the $c_n$'s as in \eqref{cncoherent} for $a \sim \ell_\textsf{NL}$: this still results in a localized particle in position space for some time. In \S\ref{decompsec}-\S\ref{nearmaxsec} we perform the general analysis for $c_n$'s chosen as in \eqref{cncoherent}, specializing to small oscillations in \S\ref{smalloscsection}. In \S\ref{generalcoherentsec} we then study other families of coherent states than those defined by \eqref{cncoherent}.

It will be useful later on to have related the oscillation amplitude $a$ (or $\alpha$) to the quantum number $n_a$ associated with the same energy, in the quadratic regime:
\begin{equation} \label{classicalquantumrelation}
    E_a = \frac{m \omega^2}{2} a^2 = \hbar \omega \alpha^2 = \hbar \omega \left( n_a + 1/2 \right) \,.
\end{equation}

We now turn to the computation of the time-dependent current \eqref{currentgeneralexpr} at $x = x_T$ in the free region, where $V(x_T) \leq 0$, at early times $t \ll t_\textsf{dec}$. We do this in two equivalent ways below (\S\ref{decompsec}-\S\ref{PIsec}). In both we first compute the wave function $\psi(x,t)$ for $x$ in the free region, and then construct the current from \eqref{currentgeneralexpr}. In this general expression for the current, we then zoom in on the time of maximal tunneling in \S\ref{nearmaxsec}.

%%%%%%%%%%%%%%%%%%%%%%%%%%%%%%%%%%%%%%%%%%%%%%%%
\subsection{Calculation by decomposition into resonant states} \label{decompsec}
%%%%%%%%%%%%%%%%%%%%%%%%%%%%%%%%%%%%%%%%%%%%%%%%
Here we express $\psi(x,t)$ as in \eqref{psiHtildeevolution}, using \eqref{psiOutgoing} for the resonant states in the free region and the semiclassical approximations of \S\ref{WKBsec}. At early times $t \ll t_\textsf{dec}$, we have
\begin{align}
    \psi(x_T,t) &= \sum_n c_n \psi_n(x_T) \, e^{-i E_n t/\hbar} e^{-\Gamma_n t/2} \notag \\
    &\approx e^{-\alpha^2/2} \sum_{n=0}^\infty \frac{\alpha^n}{\sqrt{n!}} \sqrt{\frac{m}{g_n T_n k_n(x_T)}}
    \exp \left( - \frac{S_n}{\hbar} + \frac{i\zeta_n}{\hbar} - \frac{i E_n t}{\hbar} \right) \,, \label{psixTearly}
\end{align}
and
\begin{align} \label{psidecompresult2}
    \partial_x \psi(x,t)\Big|_{x=x_T}
    &\approx \frac{i}{\hbar} e^{-\alpha^2/2} \sum_{n=0}^\infty \frac{\alpha^n}{\sqrt{n!}} \sqrt{\frac{m k_n(x_T)}{g_n T_n}}
    \exp \left( - \frac{S_n}{\hbar} + \frac{i\zeta_n}{\hbar} - \frac{i E_n t}{\hbar} \right)
\end{align}
to first subleading semiclassical order. Since the two expressions differ only in the prefactors, we can focus on \eqref{psixTearly}.

To proceed, we will approximate the sums by integrals, replacing all functions defined for integer $n$ by smooth extensions. Assuming that the dominant contributions occur at large $n$, we can set $g_n \to 1$ and use Stirling's approximation for the factorial:
\begin{align}
    \psi(x_T,t) &\approx e^{-\alpha^2/2} \int_0^\infty \di n \sqrt{\frac{m}{T_n k_n(x_T)}} \, (2\pi n)^{-1/4} e^{f(n)} \,, \\
    f(n) &= n \log \alpha-\frac{n}{2}\log n+\frac{n}{2}
    -\frac{S_n}{\hbar}+\frac{i\zeta_n}{\hbar}-\frac{i E_n t}{\hbar} \,. 
\end{align}
The integral is evaluated by saddle-point. For generic times $t$, the saddle occurs at $n_*(t) \in \mathbb{C}$. It satisfies $f'(n_*) = 0$. Using $*$ to denote evaluation at $n_*$, and defining
\begin{equation}
   \partial_n E_n = \hbar \omega_n \,, \quad \tau_n = \int_{a_n}^{b_n} \frac{m}{\kappa_n} \,, \quad t_{b_n \to x_T} = \int_{b_n}^{x_T} \frac{m}{k_n} \,, \label{omegantauntndefs}
\end{equation}
we find $n_*$ satisfies the implicit relation
\begin{equation} \label{nstarrelation}
	n_* = \alpha^2 \exp \left[ 2 \omega_* \left( \tau_* - i \left( t - t_* \right) \right) \right] \,, \quad t_* \equiv t_{b_* \to x_T} \,.
\end{equation}
Then, we have
\begin{align}
	f_* = \left( \frac{1}{2} - \omega_* \tau_* \right) n_* - \frac{S_*}{\hbar} + \frac{i \zeta_*}{\hbar} + i \omega_* n_* (t-t_*) - \frac{i E_* t}{\hbar} \,, \quad f''_* = - \frac{1}{2 n_*} - \frac{S''_*}{\hbar} + \frac{i \zeta_*''}{\hbar} - i \omega'_* t \,, \label{fppexpr}
\end{align}
where $' \equiv \partial_n$.
%MK - tried to simplify this footnote
All in all,\footnote{
The saddle-point approximation requires $|n_*|\gg1$ and a width small compared to its distance from $n=0$. When $f_*''$ is of order $1/n_*$, this width is of order $\sqrt{|n_*|}$. Thus $\alpha\gg1$ is sufficient but not necessary: $\alpha=\mathcal{O}(1)$ can also give $|n_*|\gg1$.}
%\footnote{\label{consistencyfootnote}A consistency condition on our assumption of large $n$, is that the width of the Gaussian is much smaller than its mean. If we assume the last two terms in \eqref{fppexpr} for $f_*''$ are not more important than the first, the variance is $n_*$ and the mean is $n_*$. Under this assumption, a sufficient consistency condition is, since $|n_*| \geq \alpha^2$ from \eqref{nstarrelation}, that $\alpha \gg 1$. (Here, we considered times $t$ close to the time of maximal decay, see \S\ref{nearmaxsec}.)}
\begin{align} \label{psidecompresult}
   \psi(x_T,t) \approx e^{-\alpha^2/2} \sqrt{\frac{2 \pi m}{T_* k_*(x_T) (-f''_*)}} (2\pi n_*)^{-1/4} e^{f_*} \,.
\end{align}
The current is then
\begin{grayequation} \label{Jexpressiongeneral}
    j(x_T,t) \approx \frac{e^{-\alpha^2}}{|T_* f''_*|} \sqrt{\frac{2 \pi}{|n_*|}}
    \cos \left( \arg k_*(x_T) \right)
    e^{2\,\Re f_*} \,.
\end{grayequation}

So far, the current has been expressed in terms of the implicit, complex solution $n_*(t)$ of the saddle-point equation \eqref{nstarrelation}.\footnote{To be sure, $\tau_* = \tau_{n_*(t)}$ where $\tau_n$ was defined in \eqref{omegantauntndefs} and analytically continued to $n \in \mathbb{C}$. So $\tau_n, n \in \mathbb{C}$, is a contour integral between complex WKB turning points $a_n, b_n \in \mathbb{C}$ (reminiscent of the calculation of the reflection coefficient for a wave approaching an analytic barrier with energy above the top of the barrier \cite{Berry1982,Janssen:2026ybl}). Similarly for $t_*$, which should not be confused with $T_*$: the former is the time to go from $b_*$ to $x_T$ outside of the well, while the latter is the full period inside the well defined in \eqref{classicaloscillationtime}.} We compare this formal answer with an alternative path integral computation in the next subsection. In \S\ref{nearmaxsec}, we will solve \eqref{nstarrelation} perturbatively around the time $t = t_0$ of maximal tunneling. This will still involve an implicit solution to an equation, but there it will be real and time-independent. In \S\ref{smalloscsection} we consider \textit{small} oscillations, for which we can solve \eqref{nstarrelation} explicitly for all times $t \ll t_\textsf{dec}$.

%%%%%%%%%%%%%%%%%%%%%%%%%%%%%%%%%%%%%%%%%%%%%%%%
\subsection{Calculation by semiclassical approximation to a path integral} \label{PIsec}
%%%%%%%%%%%%%%%%%%%%%%%%%%%%%%%%%%%%%%%%%%%%%%%%
Here we find $\psi(x,t)$ by semiclassically approximating the path integral that computes it \cite{feynman2010quantum}:
\begin{align}
    \psi(x,t) &= \int_\mathbb{R} \di x_0 \overset{y(t) = x}{\underset{y(0) = x_0}{\int}} [\di y(t')] \exp \left( \frac{i}{\hbar} S[y] \right) \psi(x_0,0) \\
    &= \frac{1}{( \pi d^2 )^{1/4}} \int_\mathbb{R} \di x_0 \overset{y(t) = x}{\underset{y(0) = x_0}{\int}} [\di y(t')] \exp \left( \frac{i}{\hbar} S[y] - \frac{m \omega (x_0-a)^2}{2 \hbar} \right) \,,
\end{align}
where
\begin{equation} \label{actionfunctional}
    S[y] = \int_0^t \di t' \, \left( \frac{m}{2} \dot{y}^2 - V(y) \right) \,.
\end{equation}
For simplicity, in this section, we will assume the relevant energies in the initial state sum \eqref{resexp} lie in the quadratic regime -- as in \eqref{initialstateaalpha} -- although this can be generalized \cite{exercise2026}. Therefore, the current we obtain in this section should match \eqref{Jexpressiongeneral} when setting $E_* = \hbar \omega(n_*+1/2), \omega_* = \omega, T_* = 2\pi/\omega, \omega'_* = 0$ and, at large $n_*$, $k_*(x_T) = \sqrt{2m \hbar \omega( n_* + n_T )}$ where $n_T \equiv -V(x_T)/\hbar \omega$.

The path integral is over all real-valued paths $y(t)$ that begin at $y(0)=x_0$ and reach $x$ at time $t$. To find the saddle points, we must extremize the exponent with respect to both $x_0$ and $y(t')$. The first yields a boundary condition, while the second puts us on the equations of motion:
\begin{equation} \label{complexSPeqns}
    m \ddot{z}(t') = -V'(z(t')) \,, \quad \dot{z}(0) = i \omega(z_0-a) \,, \quad z_0 \equiv z(0) \,.
\end{equation}
Here we denoted the saddle by $z(t')$ to stress that it could be complex-valued, while, at this stage, $t' \in [0,t]$ is real. In general, one must search for complex values $z_0 \in \mathbb{C}$ which are such that if the complexified classical motion starts there at $t = 0$ with momentum as in \eqref{complexSPeqns}, it reaches the given point $x \in \mathbb{R}$ at time $t \in \mathbb{R}$. The contribution from such a saddle to the wave function, to first subleading order in the semiclassical limit, reads (e.g.~\cite{Schiff:2008heb,Janssen:2026ybl}, adapted to the case at hand)
\begin{align}
        \psi(x,t) &\supset \frac{\mathcal{N}_P}{( \pi d^2 )^{1/4}} \frac{1}{\sqrt{\gamma(t)}} \exp \left( \frac{i}{\hbar} S[z] - \frac{m \omega (z_0-a)^2}{2 \hbar} \right) \quad \text{as } \hbar \to 0 \,, \label{masterformula} \\
    \text{where } \quad m \ddot{\gamma}(t') &= -V''(z) \gamma(t') \,, \quad \gamma(0) = 1 \,, \quad \dot{\gamma}(0) = i \omega \,. \label{oneloop}
\end{align}
Here $\mathcal{N}_P$ is an $(x,t)$-independent normalization that must be determined by other means \cite{Janssen:2026ybl}.

For non-quadratic systems there are infinitely many saddles, and, in general, one must use an infinite-dimensional generalization of the method of steepest descent to determine which ones contribute \cite{Witten:2010cx}. We will not attempt to solve this problem here (see \cite{Shoji:2025riv} for recent numerical progress).

Instead, we will use a trick \cite{Janssen:2026ybl}: we will solve \eqref{complexSPeqns}, and later \eqref{oneloop}, by pretending that, for given $x \in \mathbb{R}$, the endpoint $t$ lies on a judiciously chosen curve $\mathcal{C} \subset \mathbb{C}$ instead of on the positive real line. \eqref{masterformula} will then be computed for $t \in \mathcal{C}$, and, at the end of the computation, we will analytically continue $t \to \mathbb{R}$.\footnote{We will not pretend to have a proof of this procedure. What we can say is that the result matches the standard WKB calculation in the five classes of paradigmatic examples discussed in \cite{Janssen:2026ybl}. This work adds a sixth class.} Put differently, we will find an expression for the only unknown that defines a saddle point, $z_0(x,t)$, for $t \in \mathcal{C}$. We can call this $z_0(x,u)$. At the end of the computation, we will send $z_0(x,u) \to z_0(x,t)$.

\paragraph{Background} Our solution $z$ to \eqref{complexSPeqns} must reach $x$, in the free region, at a time $u \in \mathbb{C}$, which, at the end of the computation, we will send to $t \in \mathbb{R}$. To construct the solution, we will assume the starting point $z_0 \in \mathbb{C}$ lies in the quadratic region of the false vacuum. The false vacuum is centered at $x \equiv 0$, with the barrier and the true vacuum on the right side, as in Fig.~\ref{fig:potential}. There, the equations of motion are 
\begin{equation}
	\ddot{z}(t') = -\omega^2 z(t') \,.
\end{equation}
With the initial condition in \eqref{complexSPeqns}, the solutions are ellipses centered at the origin of the complex plane,
\begin{align} \label{ytpx0}
    z(t') = \left( z_0 - \frac{a}{2} \right) e^{i \omega t'} + \frac{a}{2} e^{-i \omega t'} \,,
\end{align}
with $z_0$ still to be determined. Their energy is
\begin{equation} \label{energyz0}
	E = \frac{m}{2} \dot{z}^2 + V(z) = m \omega^2 a \left(z_0 - \frac{a}{2} \right) \,.
\end{equation}

\begin{figure}[h]
\centering
\includegraphics[width=350pt]{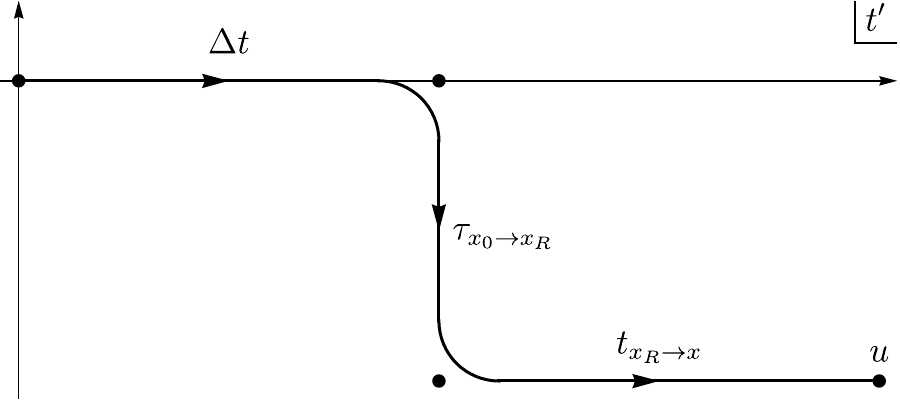}
\caption{The complex time contour taken in our solutions. The particle starts in the quadratic region of the false vacuum at $x_0 \in \mathbb{R}$, going over $k$ half periods $\Delta t = k \pi/\omega$ before reaching the point $|x_0|$ on the barrier-side of the false vacuum. Going in the negative imaginary time direction from there, the particle moves in the inverted potential, eventually reaching the point $x_R$ on the other side of the barrier (cf.~Fig.~\ref{x0greaterthana}). From there more real time evolution ensues until the real point $x$ is reached at complex time $u$. We have drawn a small complex deformation of this contour that does not change its essential properties, and avoids the WKB turning point $x_R$ as instructed in \cite{Janssen:2026ybl}.}
\label{tpcontour}
\end{figure}

To find possible values of $z_0$, we impose that the solution first goes through some real time evolution $t' : 0 \to \Delta t$, at which time it hits the positive real line ($\equiv$ on the side of the barrier). During this time the solution may be complex-valued, but $z(\Delta t) = x_r > 0$. Writing $z_0 = x_0 + i y_0$ we have
\begin{equation}
	z(t') = x_0 \cos \omega t' - y_0 \sin \omega t' + i \left[ (x_0 - a) \sin \omega t' + y_0 \cos \omega t' \right] \,,
\end{equation}
and so
\begin{equation} \label{xtequation}
	x_r = \text{Re} \, z_0 e^{i \omega \Delta t} \,, \quad (x_0 - a) \sin \omega \Delta t + y_0 \cos \omega \Delta t = 0 \,.
\end{equation}
Then, we go through some evolution in the negative imaginary time direction. We would like the solution to be real-valued along this line.\footnote{If not, we would lose analytic control for general barriers \cite{Janssen:2026ybl}. This key assumption will be justified only at the end of the computation when we demonstrate it matches the decomposition calculation of \S\ref{decompsec}.} To achieve this, the (real) time derivative of the solution must be purely imaginary when the particle reaches $x_r$. This imposes
\begin{equation} \label{imaginaryderivativecondition}
	x_0 \sin \omega \Delta t + y_0 \cos \omega \Delta t = 0 \,.
\end{equation}
Together with \eqref{xtequation} this implies $a \sin \omega\Delta t = 0$, and since\footnote{This step is qualitatively different from the $a = 0$ calculation, where \eqref{imaginaryderivativecondition} does not introduce an independent condition: the trajectories for $a = 0$ are circles, instead of ellipses, and the derivative is always purely imaginary when we cross the positive real line. In that case we can have a continuous initial real time evolution $\omega \Delta t$ \cite{Janssen:2026ybl}.} $a > 0$ we must have
\begin{equation} \label{restrictionDeltat}
	\omega \Delta t = k \pi \,, \quad k \in \mathbb{N} \,, \quad \text{and} \quad y_0 = 0 \,.
\end{equation}
We still have a free parameter: $x_0 \in \mathbb{R}$. For $x_0 > 0$, we must take $k$ even and for $x_0 < 0$ we must take $k$ odd to have $x_r = x_0 \cos \omega \Delta t = (-1)^k x_0 = |x_0| > 0$, ending up on the barrier side of the false vacuum after time $\Delta t$. We have
\begin{align} \label{energyx0}
	\dot{z}(\Delta t) = (-1)^k i \omega (x_0 - a) \,, \quad E(x_0) &= m \omega^2 a \left( x_0 - \frac{a}{2} \right) = V(x_0) - V(x_0-a) \,.
\end{align}
Writing $t' = \Delta t - i \tau$ and $' \equiv \di/\di \tau$, we have in the negative imaginary time direction
\begin{equation}
	z(\tau = 0) = |x_0| \,, \quad z'(\tau =0) = \omega(|x_0| - (-1)^k a) \,, \quad \frac{m}{2} (z')^2 - V(z) = - E(x_0) \,,
\end{equation}
where the advantage is that we can use the last equation also outside of the quadratic region without losing analytic control; the particle remains on the real line, moving in the inverted potential $-V$ at energy $-E$. Finally, we would like the particle to come to a halt on the other side of the barrier. This is the only way for the particle to remain real-valued once we make a final 90-degree turn in the real time direction, after which the particle rolls down the upright potential to reach $x$. This contour in the complex $t'$-plane is shown in Fig.~\ref{tpcontour}.

\begin{figure}[h]
\centering
\includegraphics[width=400pt]{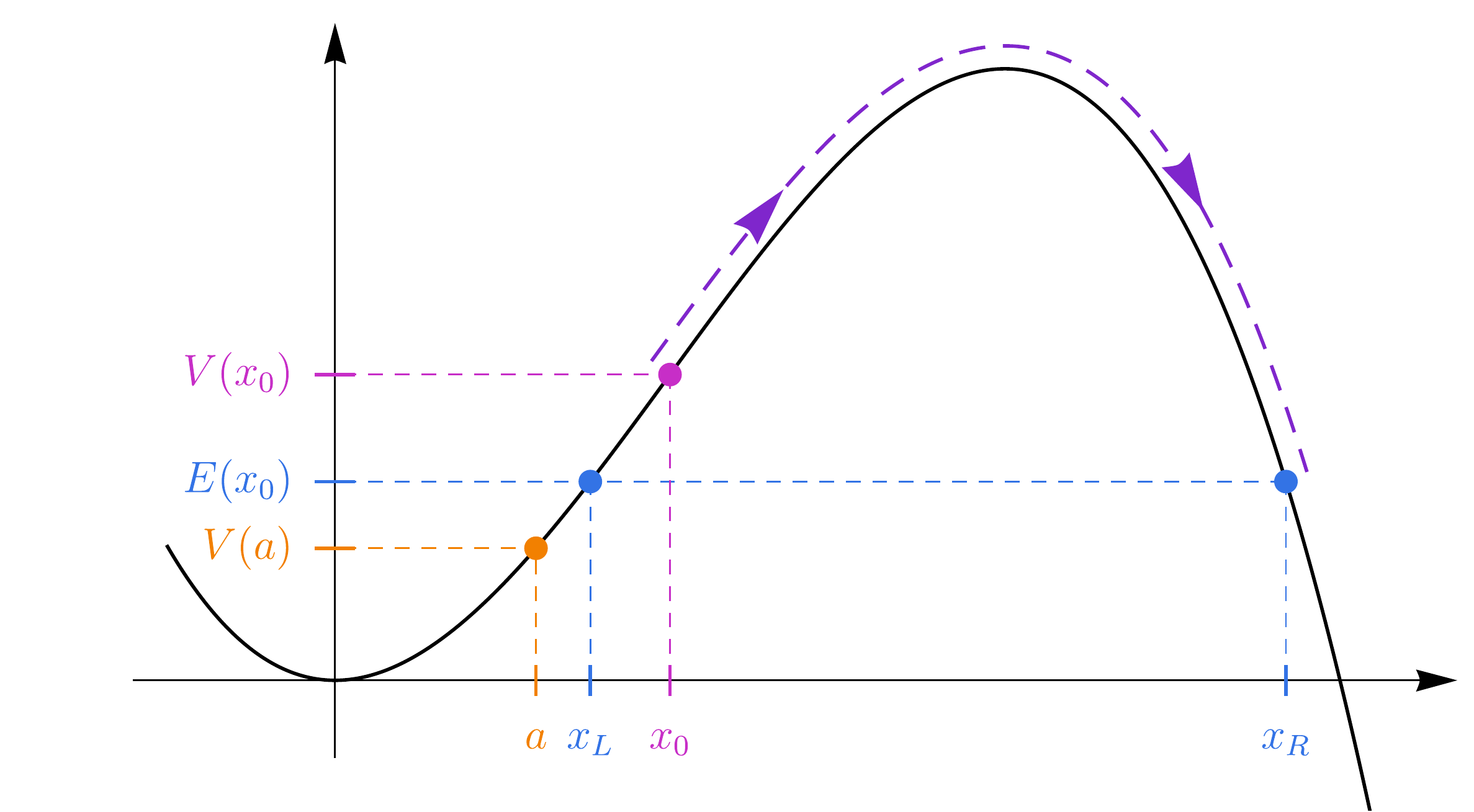}
\caption{A sketch of the tunneling solution for $x_0 > a$, described around \eqref{energyintervalx0greaterthana}. The particle starts at $x_0$ and first goes over $k/2 \in \mathbb{N}$ full (complex, elliptical) oscillations in the quadratic false vacuum region, returning to $x_0$ (not shown). Then time turns in the negative imaginary direction, and the particle starts rolling over the barrier with a non-vanishing initial velocity. It eventually comes to a halt on the other side of the barrier at $x_R$ where $V(x_R) = E(x_0)$. After that, real time evolution follows until the final point $x$ to the right is reached (not shown). The particle can have a larger potential energy $V(x_0)$ than total energy $E(x_0) = V(x_L) = V(x_R)$ because the initial velocity is purely imaginary, so the kinetic energy (in real time) is negative. This picture is more intuitive in the inverted potential $-V$ at energy $-E$; there, the potential energy $-V(x_0) < -E(x_0)$ and the kinetic energy (in imaginary time) is positive.}
\label{x0greaterthana}
\end{figure}

Let us study this solution for $x_0 > a$ (see Fig.~\ref{x0greaterthana}).\footnote{This continuous range of $x_0$ will suffice for our purpose of analytically continuing the result. We will not discuss the range $x_0 \leq a$.} In particular $x_0 > 0$ so $k$ is even, and $z'(0) = \omega(x_0 - a) > 0$. Also
\begin{equation} \label{energyintervalx0greaterthana}
	V(a) < E(x_0) < V(x_0) \,.
\end{equation}
(This last, unusual inequality is possible because the velocity in the real time direction is purely imaginary, so the kinetic energy is negative.) So during the imaginary time evolution, our particle starts with some positive velocity in the inverted potential, and comes to rest at a certain point $x_R$ on the other side of the barrier where 
\begin{equation} \label{bdefinition}
	E(x_0) = V(x_R) \,.
\end{equation}	
This point lies after the equipotential point corresponding to $x_0$ but before the one corresponding to $a$. The amount of imaginary time spent during this transition is
\begin{equation}
	\tau_{x_0 \to x_R} = \int_{x_0}^{x_R} \frac{m}{\kappa_{x_0}} \,, \quad \kappa_{x_0}(x') = \sqrt{2m \left( V(x') - E(x_0) \right)} \,,
\end{equation}
and the remaining real time on the other side is
\begin{equation} \label{realtimeeq}
	t_{x_R \to x} = \int_{x_R}^x \frac{m}{k_{x_0}} \,, \quad k_{x_0}(x') = \sqrt{2m \left( E(x_0) - V(x') \right)} \,.
\end{equation}

Now, consider a fixed even $k$, and vary $x_0 > a$. This traces out a \textit{curve} in the complex $u$-plane of possible end points of the solution:
\begin{equation} \label{uvsx0bar}
	u = \frac{k \pi}{\omega} - i \tau_{x_0 \to x_R} + t_{x_R \to x} \,, \quad k \in 2\mathbb{N} \,.
\end{equation}
(Recall $x_R$ depends on $x_0$, via $V(x_R) = E(x_0)$.) Therefore we can imagine doing the computation for various values of the continuous parameter $x_0 > a$, and then, at the end of the computation, analytically continuing $x_0(u) \to x_0(t)$ via the inverse of the relation \eqref{uvsx0bar}.

\paragraph{On-shell exponent} A straightforward calculation yields for the on-shell exponent
\begin{align} \label{OSexponent}
	\mathfrak{f} = i S[z] - \frac{m \omega}{2} (z_0-a)^2 = - S_{x_0 \to x_R} + i \phi_{x_R \to x} + i E(x_0) \left( \frac{k \pi}{\omega} - u \right) - \frac{m \omega}{2} (x_0-a)^2 \,,
\end{align}
where
\begin{equation}
	S_{x_0 \to x_R} = \int_{x_0}^{x_R} \kappa_{x_0} \,, \quad \phi_{x_R \to x} = \int_{x_R}^x k_{x_0} \,.
\end{equation}
Now in this result we must analytically continue $x_0(u) \to x_0(t)$ via \eqref{uvsx0bar}.

To do this, it is easier to trade $x_0$ for $x_L$, the equipotential point corresponding to $x_R$. It satisfies $V(x_L) = E(x_0)$ on the false vacuum side -- assumed to be in the quadratic regime -- so
\begin{equation} \label{x0xLrelation}
	x_0 = \frac{x_L^2 + a^2}{2a} \,.	
\end{equation}
Since $x_0 > x_L > a$, instead of $x_0$ we can vary $x_L > a$ continuously. In the defining equation \eqref{uvsx0bar} for $x_0(u)$, we add and subtract a piece $\tau_{x_L \to x_0}$ to $\tau_{x_0 \to x_R}$, and use \eqref{x0xLrelation} elsewhere so that only $x_L$ remains. We find
\begin{equation}
	\omega \tau_{x_L \to x_0} = \log \frac{x_L}{a} \,.
\end{equation}
Therefore
\begin{equation}
	\omega \left( u - t_{x_R \to x} - \frac{k \pi}{\omega} + i \tau_{x_L \to x_R} \right) = i \log \frac{x_L}{a} \,,
\end{equation}
 or, in other words, $x_L$ satisfies
\begin{equation} \label{definingxL}
	x_L = a \exp \left( \omega \tau_{x_L \to x_R} - i \omega \left( u - t_{x_R \to x} \right) \right) \,,
\end{equation}
where we used that $e^{i \pi k} = 1$ for even $k$. So $k$ drops out. Instead of summing over all $k$, which would give an infinite answer, we will assume that at any given time a single solution with some $k$ contributes. This makes sense because $k$ corresponds to the amount of oscillations that have been made in the well.

From \eqref{classicalquantumrelation}, a classical oscillation amplitude $A$ corresponds to a quantum excitation number
\begin{equation} \label{quantumclassicalcorrespondence}
	n_A + 1/2 = \alpha_A^2 = \frac{A^2}{2 d^2} \,.
\end{equation}
Therefore, we have, squaring \eqref{definingxL} and analytically continuing $u \to t$,
\begin{equation} \label{nxLequation}
	n_{x_L} + 1/2 = \alpha_a^2 \exp \left[ 2 \omega \left( \tau_{x_L \to x_R} - i \left( t - t_{x_R \to x} \right) \right) \right] \,,
\end{equation}
where $t_{x_R \to x}, \tau_{x_L \to x_R}$ are to be computed at energies $E(n_{x_L}) = \hbar \omega (n_{x_L}+1/2) = m \omega^2 x_L^2/2 = V(x_L) = V(x_R)$. This is \textit{almost} the equation we found for the saddle point $n_*(t)$ in Eq.~\eqref{nstarrelation}, identifying $\alpha_a \cong \alpha$, except for the $1/2$ on the left-hand side of \eqref{nxLequation}. The solutions are related by
\begin{equation}
    n_{x_L} \approx n_* + \delta \,, \quad \delta = \frac{1}{4 \omega n_*(\tau_*'+i t'_*) - 2} \,.
\end{equation}
Therefore, to translate to the language of excitation numbers used in \S\ref{decompsec}, we should identify
\begin{equation} \label{nstarxLrelation}
	n_* + \frac{1}{2} + \delta = \frac{x_L^2}{2 d^2} \,, \quad \alpha^2 = \frac{a^2}{2 d^2} \,.
\end{equation}

Now, the on-shell exponent in \eqref{OSexponent} becomes, in terms of $x_L$,
\begin{equation}
	\frac{\mathfrak{f}}{\hbar} = - \frac{a^2}{4 d^2} + \left( \frac{1}{2} - \omega \tau_{x_L \to x_R} \right) \frac{x_L^2}{2d^2} -\frac{S_{x_L \to x_R}}{\hbar} + \frac{i}{\hbar} \left( \phi_{x_R \to x} - V(x_L) t_{x_R \to x} \right) \,.
\end{equation}
Using the dictionary \eqref{nstarxLrelation} and dropping terms beyond the next-to-leading order in the semiclassical limit, we find
\begin{align}
	\frac{\mathfrak{f}}{\hbar} &= - \frac{\alpha^2}{2} - \frac{i \pi}{4} - \frac{\omega \tau_*}{2} + \frac{i \omega (t-t_*)}{2} + f_* \,, \label{OSresult}
\end{align}
with $f_*$ as in \eqref{fppexpr}. (Here, we have not yet specialized to $x = x_T$, so in the definition \eqref{AnphaseSC} of $\zeta$, we should remember that here we have $x_T \to x$.) Therefore this procedure reproduces the semiclassical exponent $f_*$ we obtained via the decomposition method, plus a residual $\mathcal{O}(\hbar^0)$ piece, in the case for which the two computations should agree discussed below \eqref{actionfunctional}. We now turn to the prefactor.

\paragraph{Prefactor} We must solve \eqref{oneloop}, with $z(t')$ the background we constructed above. Recall that $t': 0 \to u \in \mathbb{C}$ following the contour as in Fig.~\ref{tpcontour}, and we will analytically continue $u \to t$ only at the end of the computation. So, initially, $z(t')$ is located in the quadratic region of the false vacuum, where $V''(z) = m \omega^2$. The equation for $\gamma$ becomes $\ddot{\gamma} = - \omega^2 \gamma$ and so the solution in this region is
\begin{equation}
	\gamma(t') = e^{i \omega t'} \,.
\end{equation}
To continue this solution in the negative imaginary time direction, where eventually the potential is no longer quadratic, it is useful to express it in terms of $z(t')$. This can be done by noting that the following two functions are linearly independent solutions to $m \ddot{\gamma}(t') = -V''(z) \gamma(t')$ (e.g.~\cite{PhysRevD.25.330}):
\begin{align}
    \gamma_1(t') = \dot{z}(t') \,, \quad \gamma_2(t') = \dot{z}(t') \int_0^{t'} \frac{\di t''}{\dot{z}(t'')^2} \,.
\end{align}
In the quadratic region, they read
\begin{equation}
    \gamma_1(t') = i \omega \left( x_0 \, e^{i \omega t'} - a \cos \omega t' \right) \,, \quad \gamma_2(t') = - \frac{i \sin \omega t'}{\omega^2 (x_0 - a)} \,.
\end{equation}
Therefore, we can write
\begin{equation}
    \gamma(t') = c_1 \gamma_1(t') + c_2 \gamma_2(t') \,, \quad c_1 = \frac{1}{i \omega(x_0-a)} \,, \quad c_2 = a \omega^2 \,.
\end{equation}
Next, we must evaluate $\gamma(t')$ at $t' = u$. Our classical particle conserves its energy $E(x_0)$ given in \eqref{energyx0}, so
\begin{equation} \label{momentumatu}
    m \dot{z}(u) = k_{x_0}(x)
\end{equation}
with the momentum given in \eqref{realtimeeq}. We will be interested in evaluating this at $x = x_T$ in the free region, therefore\footnote{When we compute the current by taking an $x$-derivative, to leading order in the semiclassical limit this derivative acts on the exponent, so we will not need to take a derivative of \eqref{momentumatu}.}
\begin{equation}
    m \dot{z}(u)
    = \sqrt{2m \left( E(x_0)-V(x_T) \right)}
    = \sqrt{2m \left( V(x_L)-V(x_T) \right)}
    = m \omega \sqrt{x_L^2+2d^2 n_T} \,.
\end{equation}
Then, we have
\begin{align}
    \int_0^u \frac{\di t'}{\dot{z}(t')^2} &= \int_0^{\Delta t} \frac{\di t'}{\dot{z}(t')^2} + i \underset{\tau_{x_0 \to x_R}}{\int} \frac{\di \tau'}{z'(\tau')^2} + \underset{t_{x_R \to x_T}}{\int} \frac{\di t'}{\dot{z}(t')^2} \\
    &= 0 + i \int_{x_0}^{x_R} \frac{m^3}{\kappa_{x_0}^3} + \int_{x_R}^{x_T} \frac{m^3}{k_{x_0}^3} \,, \label{intermediateintegral}
\end{align}
where we have again used the conservation of energy. Naively, the two integrals we have written have a divergence from near $x_R$. However, in Fig.~\ref{tpcontour}, we avoided $x_R$ by passing around it \textit{from above} in the complex plane. So we should interpret the integrals in \eqref{intermediateintegral} as being done along contours slightly displaced into the upper-half plane, making them finite. Now, keeping this deformation in mind, we can re-express the integrals as
contour integrals around branch cuts, as follows\footnote{We thank Barak
Gabai for this suggestion.}:
\begin{align}
    i \int_{x_0}^{x_R} \frac{m^3}{\kappa_{x_0}^3} + \int_{x_R}^{x_T} \frac{m^3}{k_{x_0}^3}
    \;&=\;
    \begin{tikzpicture}[cdraw]
        \node at (2,0.82) {\small$im^{3}\!\int\!\big(2m(V-V_R)\big)^{-3/2}$};
        % integration contour (real axis): x_0 -> over x_R -> x_T
        \draw[flow] (0,0.13) -- (1.78,0.13);
        \draw (1.78,0.13) arc (180:0:0.22);
        \draw[flow] (2.22,0.13) -- (4,0.13);
        % branch point at x_R, sitting at the end of the cut, which runs left
        \draw[bcut] (2,0) -- (0,0);
        \node at (2,0) {$\times$};
        % x_0 and x_T marked as dots on the real line
        \fill (0,0) circle (1.1pt);
        \fill (4,0) circle (1.1pt);
        \node[below] at (0,0) {$x_0$};
        \node[below] at (4,0) {$x_T$};
        \node[below] at (2,0) {$x_R$};
    \end{tikzpicture} \\[1.6ex]
    \;&=\;
    \frac{1}{2}\;
    \begin{tikzpicture}[cdraw]
        \node at (1.75,0.82) {\small$im^{3}\!\int\!\big(2m(V-V_R)\big)^{-3/2}$};
        % half-contour wrapping the cut on [x_0,x_R], branch point at right
        \draw[flow] (0,0.20) -- (3.4,0.20);
        \draw (3.4,0.20) arc (90:-90:0.20);
        \draw[flow] (3.4,-0.20) -- (0,-0.20);
        \draw[bcut] (3.3,0) -- (0,0);
        \node at (3.3,0) {$\times$};
        % x_0 marked as a dot on the real line (left tip of the cut)
        \fill (0,0) circle (1.1pt);
        \node[below] at (0,-0.20) {$x_0$};
        \node[below] at (3.4,-0.20) {$x_R$};
    \end{tikzpicture}
    \;+\;
    \frac{1}{2}\;
    \begin{tikzpicture}[cdraw]
        \node at (1.85,0.82) {\small$m^{3}\!\int\!\big(2m(V_R-V)\big)^{-3/2}$};
        % half-contour wrapping the cut on [x_R,x_T], branch point at left
        \draw[flow] (0.2,0.20) -- (3.6,0.20);
        \draw (0.2,0.20) arc (90:270:0.20);
        \draw[flow] (3.6,-0.20) -- (0.2,-0.20);
        \draw[bcut] (0.3,0) -- (3.6,0);
        \node at (0.3,0) {$\times$};
        % x_T marked as a dot on the real line (right tip of the cut)
        \fill (3.6,0) circle (1.1pt);
        \node[below] at (0.3,-0.20) {$x_R$};
        \node[below] at (3.6,-0.20) {$x_T$};
    \end{tikzpicture}
    \,.
\end{align}
We can extend the first integral to encircle the branch cut between
$x_L < x_0$ and $x_R$, plus a correction which we can compute assuming that
$[x_L,x_0]$ lies in the quadratic region:
\begin{align}
    \begin{tikzpicture}[cdraw]
        \node at (1.75,0.82) {\small$im^{3}\!\int\!\big(2m(V-V_R)\big)^{-3/2}$};
        % half-contour wrapping the cut on [x_0,x_R], branch point at right
        \draw[flow] (0,0.20) -- (3.4,0.20);
        \draw (3.4,0.20) arc (90:-90:0.20);
        \draw[flow] (3.4,-0.20) -- (0,-0.20);
        \draw[bcut] (3.3,0) -- (0,0);
        \node at (3.3,0) {$\times$};
        % x_0 marked as a dot on the real line
        \fill (0,0) circle (1.1pt);
        \node[below] at (0,-0.20) {$x_0$};
        \node[below] at (3.4,-0.20) {$x_R$};
    \end{tikzpicture}
    \;&=\;
    \begin{tikzpicture}[cdraw]
        \node at (1.85,0.82) {\small$im^{3}\!\int\!\big(2m(V-V_R)\big)^{-3/2}$};
        % closed contour encircling the cut on [x_L,x_R], branch points at both ends
        \draw[flow] (0.2,0.20) -- (3.4,0.20);
        \draw (3.4,0.20) arc (90:-90:0.20);
        \draw[flow] (3.4,-0.20) -- (0.2,-0.20);
        \draw (0.2,-0.20) arc (270:90:0.20);
        \draw[bcut] (0.3,0) -- (3.3,0);
        \node at (0.3,0) {$\times$};
        \node at (3.3,0) {$\times$};
        % x_0 marked as a dot on the real line, between x_L and x_R
        \fill (1.3,0) circle (1.1pt);
        \node[below] at (0.3,-0.20) {$x_L$};
        \node[below] at (1.3,-0.20) {$x_0$};
        \node[below] at (3.4,-0.20) {$x_R$};
    \end{tikzpicture}
    \;-\;
    \begin{tikzpicture}[cdraw]
        \node at (0.8,0.82) {\small$i \omega^{-3}\!\int\!\big(x^{2}-x_L^{2}\big)^{-3/2}$};
        % short half-contour wrapping the cut on [x_L,x_0], branch point at left.
        % spans the same x_L -> x_0 distance as inside the [x_L,x_R] contour above
        \draw[flow] (0.2,0.20) -- (1.3,0.20);
        \draw (0.2,0.20) arc (90:270:0.20);
        \draw[flow] (1.3,-0.20) -- (0.2,-0.20);
        \draw[bcut] (0.3,0) -- (1.3,0);
        \node at (0.3,0) {$\times$};
        % x_0 marked as a dot on the real line (right tip of the cut)
        \fill (1.3,0) circle (1.1pt);
        \node[below] at (0.3,-0.20) {$x_L$};
        \node[below] at (1.3,-0.20) {$x_0$};
    \end{tikzpicture} \\[1.6ex]
    \;&=\;
    \begin{tikzpicture}[cdraw]
        \node at (1.85,0.82) {\small$im^{3}\!\int\!\big(2m(V-V_R)\big)^{-3/2}$};
        % closed contour encircling the cut on [x_L,x_R], branch points at both ends
        \draw[flow] (0.2,0.20) -- (3.4,0.20);
        \draw (3.4,0.20) arc (90:-90:0.20);
        \draw[flow] (3.4,-0.20) -- (0.2,-0.20);
        \draw (0.2,-0.20) arc (270:90:0.20);
        \draw[bcut] (0.3,0) -- (3.3,0);
        \node at (0.3,0) {$\times$};
        \node at (3.3,0) {$\times$};
        % x_0 marked as a dot on the real line, between x_L and x_R
        \fill (1.3,0) circle (1.1pt);
        \node[below] at (0.3,-0.20) {$x_L$};
        \node[below] at (1.3,-0.20) {$x_0$};
        \node[below] at (3.4,-0.20) {$x_R$};
    \end{tikzpicture}
    \;+\;
   \frac{2 i x_0}{\omega^3 x_L^{2}\sqrt{x_0^{2}-x_L^{2}}} \,.
\end{align}

In the same way, we can express
\begin{align}
    -S_n + i \phi_n &= -(1/2) \underset{[a_n,b_n]}{\ointclockwise} \kappa_n + (i/2) \hspace{-0.5cm} \underset{x_T - i \varepsilon \to b_n \to x_T + i \varepsilon}{\int} \hspace{-0.8cm} k_n \,,
\end{align}
and so
\begin{align}
    -S_n'' + i \phi_n'' = -S''_n + i \zeta_n'' &= -i \frac{\hbar^2 \omega^2}{m} \left( i \int_{x_0}^{x_R} \frac{m^3}{\kappa_{x_0}^3} + \int_{x_R}^{x_T} \frac{m^3}{k_{x_0}^3} - \frac{i x_0}{\omega^3 x_L^2 \sqrt{x_0^2 - x_L^2}} \right) \\
    &= -i \frac{\hbar^2 \omega^2}{m} \left( \int_0^u \frac{\di t'}{\dot{z}(t')^2} - \frac{i (x_L^2 + a^2)}{\omega^3 x_L^2 (x_L^2 - a^2)} \right) \,.
\end{align}
In the second equality, we moved the derivatives $'' \equiv \partial_n^2$ through the integral, which is justified because by Cauchy's theorem we can have the encircling contours be independent of $n$. This, and the related manifest absence of divergences, is the usefulness of re-expressing the action integrals as contour integrals. We have also used the correspondence between excitation numbers $n$ and amplitudes repeated around \eqref{quantumclassicalcorrespondence} and the relation \eqref{x0xLrelation} between $x_0$ and $x_L$.

Putting these results together, we find
\begin{align}
    \gamma(u)
    = -\frac{ia\sqrt{x_L^2+2d^2 n_T}}{x_L^2}
    \left[
        1-2n_*\left(
            -\frac{S''_*}{\hbar}+\frac{i\zeta''_*}{\hbar}
        \right)
    \right] \,,
\end{align}
where we have used \eqref{nstarxLrelation}. Inserting this in \eqref{masterformula} with\footnote{That this is the correct normalization for this instanton follows by analogy with the scattering example in \cite{Janssen:2026ybl}.} $\mathcal{N}_P = 1$, using \eqref{definingxL}-\eqref{OSresult} and sending $u \to t$, we find
\begin{align} \label{psiPIresult}
    \psi(x_T,t)
    &= \frac{e^{-\alpha^2/2}}{(\pi d^2)^{1/4}}
    \left( \frac{n_*}{n_*+n_T} \right)^{1/4}
    \frac{e^{f_*}}{
        \sqrt{1-2n_*\left(-S_*''+i\zeta_*''\right)/\hbar}
    } \,.
\end{align}
This matches the decomposition result \eqref{psidecompresult}, for the case discussed below \eqref{actionfunctional}. Using the
relation obtained by taking $\partial_x$ of \eqref{definingxL},
we find that
$\left.\partial_x \left( \mathfrak{f}/\hbar \right)\right|_{x=x_T}
= i k_n(x_T)/\hbar
= i m\omega\sqrt{x_L^2+2d^2 n_T}/\hbar
\to i\sqrt{2\hbar m\omega(n_*+n_T)}/\hbar$,
using \eqref{nstarxLrelation} at large $n_*$.
To leading order in the semiclassical limit, the derivative acts
on the exponent, so multiplying \eqref{psiPIresult} by this factor
reproduces \eqref{psidecompresult2}.

With this we have shown that the path integral computation of the wave function and its derivative at $x_T$, and hence the current, agree with the decomposition method of \S\ref{decompsec}. Both yield the result \eqref{Jexpressiongeneral}, valid for all $t \ll t_\textsf{dec}$. We now zoom in on the time of maximal current.

%%%%%%%%%%%%%%%%%%%%%%%%%%%%%%%%%%%%%%%%%%%%%%%%%%%%%%%%%%%%%%%%%%%%%%%%%%%%%%%%%%%%%%%%%%%%%%%%
\subsection{Perturbation theory near the time of maximal decay} \label{nearmaxsec}
%%%%%%%%%%%%%%%%%%%%%%%%%%%%%%%%%%%%%%%%%%%%%%%%%%%%%%%%%%%%%%%%%%%%%%%%%%%%%%%%%%%%%%%%%%%%%%%%
Consider the saddle-point equation \eqref{nstarrelation} for $n_*(t)$, setting $t = t_*$ and denoting $n_*(t_*) \equiv n_0$:
\begin{equation} \label{n0equation}
    n_0 = \alpha^2 e^{2 \omega_0 \tau_0} \,.
\end{equation}
This presumably has a real solution $n_0$, which we assume has been found. Therefore our assumption of setting $t$ equal to $t_* \to t_0 = t_{b_{n_0} \to x_T}$ was consistent; in general, $t_*$ is a complex time, so we cannot set $t = t_*$ during the actual real time evolution of the current, but we see there is a special time $t_0$ for which it is real. This is the time at which the current at $x_T$ is maximal. The time delay between $t = 0$ (when the particle is at its maximum amplitude in the well, Eq.~\eqref{initialstateaalpha}) and $t = t_0$ is expected because, while the imaginary time transition takes no real time, it takes the particle some time $t_0$ to roll from $b_{n_0}$ to $x_T$ after it has emerged on the other side of the barrier. The solution to \eqref{n0equation} determines this time for the oscillating particle in the well.

Recall from Eq.~\eqref{classicalquantumrelation} the relation $n \sim \alpha^2$ between the quantum excitation number and the rescaled classical amplitude of the particle. The saddle point $n_0$ satisfying \eqref{n0equation} has $n_0 > \alpha^2$. This can be understood as coming from the finite width of the initial state when the particle is at its maximal height in the well: it is favorable for the particle to ``start tunneling" from higher up in the potential, because there is less barrier to traverse, but on the other hand this is penalized by the wave function. The balance of these two effects results in \eqref{n0equation}. Decays are weighted by $|c_n|^2\Gamma_n$, rather than $|c_n|^2$ only, and can therefore have energy $E(n_0)$ much greater than the mean. Repeating the decay experiment with the same state preserves this bias. Only when  detections at $x_T$ at all times are included, including very late times after the state has changed significantly, would the energy histogram of decaying particles reproduce $|c_n|^2$.\footnote{We thank Mehrdad Mirbabayi for conversations on this point.}

After these preliminary remarks, we study the current to first subleading order in perturbation theory around $t = t_0$. To this end we expand the solution to \eqref{nstarrelation} as
\begin{equation} \label{nstarexpansion}
	n_*(t) = n_0 \left( 1 + n_1 (t-t_0) + \cdots \right) \,,
\end{equation}
valid for $|n_1 (t-t_0)| \ll 1$. Then we find
\begin{equation} \label{n1sol}
	n_1 = -\frac{2 i \omega_0}{1 - 2 n_0 \left[ \omega_0 \left( \tau'_0 + i t'_0 \right) + \omega'_0 \tau_0 \right]} \,.
\end{equation}
If we can neglect the derivative terms in the denominator, the consistency condition of the expansion \eqref{nstarexpansion} is simply $|\omega_0 (t-t_0)| \ll 1$.

We do not need to go to second order in $n_*(t)$ to find the second order behavior in $t-t_0$ of $j(t)$: we find
\begin{equation}
	f_* = f(n_*(t)) = f_0 - \frac{i E_0}{\hbar}(t-t_0) - \frac{1}{2} f''_0 \left( n_0 n_1 \right)^2 (t-t_0)^2 + \mathcal{O}((t-t_0)^3) \quad \text{as } t \to t_0 \,.
\end{equation}
Neglecting the time-dependence of the prefactor in \eqref{Jexpressiongeneral}, and using $T_0 = 2\pi/\omega_0$, we have
\begin{align}
	j &\approx e^{-\alpha^2} \frac{\omega_0}{|f''_0| \sqrt{2 \pi n_0}} e^{2 \, \Re f_0} \exp \left( - \Re \left( f''_0 n_1^2 \right) n_0^2 (t-t_0)^2 \right) \,, \label{Jfinaltime} \\
	\Re \, f_0 &= \left( \frac{1}{2} - \omega_0 \tau_0 \right) n_0 - \frac{S_0}{\hbar} \,, \quad f''_0 = - \frac{1}{2n_0} + \omega_0 \left( \tau'_0 + i t'_0 \right) + \omega'_0 \tau_0 \,, \\ f''_0 n_1^2 &= \frac{2 \omega_0^2/n_0}{1 - 2 n_0 \left[ \omega_0 \left( \tau'_0 + i t'_0 \right) + \omega'_0 \tau_0 \right]} \,. \label{fpp0eq}
\end{align}
The effective time window of tunneling is
\begin{equation}
	\Delta t \sim \frac{1}{n_0 \sqrt{\Re \left( f''_0 n_1^2 \right)}} \,.
\end{equation}
If we assume the various terms in $n_1$ and $f_0''$ are of the same order of magnitude (they scale the same way with $\hbar$), we get
\begin{equation}
	\omega_0 \Delta t \sim \frac{1}{\sqrt{n_0}} = \frac{1}{\alpha \, e^{\omega_0 \tau_0}} \,,
\end{equation}
which is much smaller than one since $n_0 \gg 1$. Therefore the bulk of the tunneling happens in a short time interval compared to the classical oscillation time. The total probability that flows through the barrier during this tunneling burst is
\begin{align}\label{deltaP}
	\Delta P &= e^{-\alpha^2} \frac{\omega_0}{\sqrt{2 |f''_0|^2 \, \Re \left( f''_0 n_1^2 \right) (n_0)^3}} \exp \left( (1-2\omega_0 \tau_0)n_0 - \frac{2 S_0}{\hbar} \right) \,.
\end{align}

\paragraph{Quasistatic limit} If we are in a situation where
\begin{equation}
    \left| n_0 \left[ \omega_0 \left( \tau'_0 + i t'_0 \right) + \omega'_0 \tau_0 \right] \right| \ll 1 \,,
\end{equation}
then $f_0'' \approx -1/(2n_0), n_1 \approx -2i \omega_0$ and we have
\begin{equation}
	\frac{\omega_0}{\sqrt{2 |f''_0|^2 \, \Re \left( f''_0 n_1^2 \right) (n_0)^3}} \quad \to \quad 1 \,.
\end{equation}
If additionally the classical oscillation time in the false vacuum is much longer than the oscillation time over the barrier in the inverted potential,
\begin{equation}
    \omega_0 \tau_0 \ll 1 \,,
\end{equation}
then
\begin{equation}
    \Delta P = 1 \times \exp \left( - \frac{2S_{\alpha^2}}{\hbar} + 2(\alpha \omega_{\alpha^2} \tau_{\alpha^2})^2 + \cdots \right) \,.
\end{equation}
One might have guessed the leading $S_{\alpha^2}$ term by using a ``quasistatic" approximation for the particle near the time of maximal tunneling. We see here that -- as expected -- there is a correction coming from the finite width of the wave function, involving the change in the action over the width $\Delta n = \alpha$ of the coherent state:
$|S'_{\alpha^2}| \Delta n/\hbar = \alpha \omega_{\alpha^2} \tau_{\alpha^2}$. The correction is positive because the higher-energy components of the initial state are less suppressed by the barrier.

Under these assumptions the current near the maximum is approximately
\begin{equation} \label{jactualquasistatic}
    j \approx \sqrt{\frac{2 \alpha^2 \omega_0^2}{\pi}} e^{-2 \alpha^2 \omega_0^2 (t-t_0)^2} \Delta P \,.
\end{equation}
The effective timescale of tunneling $\Delta t \sim 1/(\alpha \omega_0)$ can be estimated in this case by demanding that the phase $\omega_0 \Delta n \Delta t$ changes by an $\mathcal{O}(1)$ amount over the width of the coherent state, causing destructive interference between the terms in the sum.
%These results agree with the results for the transmission of a wave with energy $E$ below the top of a general barrier discussed in \cite{Janssen:2026ybl}, with the identifications $E = \hbar \omega \alpha^2$ and setting the width of the incoming state to $d$.

Notice that it would \textit{not} have been correct to estimate the leading time-dependence in $j$, in the quasistatic limit, by
\begin{equation} \label{jguess}
    j \neq e^{-2 S(\langle x \rangle_t)/\hbar} \,,
\end{equation}
where $\langle x \rangle_t$ is the classical particle motion. The expansion of \eqref{jguess} to order $(t-t_0)^2$ involves $\tau_{\alpha^2}$, a property of the barrier, while in \eqref{jactualquasistatic} only the combination $\alpha \omega_0$ appears.

%%%%%%%%%%%%%%%%%%%%%%%%%%%%%%%%%%%%%%%%%%%%%%%%
\subsection{Small oscillations} \label{smalloscsection}
%%%%%%%%%%%%%%%%%%%%%%%%%%%%%%%%%%%%%%%%%%%%%%%%
In the previous subsections we made no assumption about the amplitude of oscillations in the well, apart from the requirement that the saddle $n_0$ in \eqref{n0equation} should correspond to an energy $E(n_0)$ much below the top of the barrier. In this subsection, we will assume the saddle point only explores the quadratic regime of the potential around the metastable minimum, where the oscillation amplitude scales as $a \sim d \alpha$ as explained around \eqref{initialstateaalpha}.

More specifically, we will study the regime
\begin{equation}
    d \ll \sqrt{ab} \, e^{A/2} \ll |\ell_1|, |\ell_2| \,, \label{smalloscregime}
\end{equation}
where $b$ is the point where $V(b) = V(0) = 0$ on the free side of the barrier,
\begin{equation} \label{Adefinition}
    A = \int_0^b \di x' \left( \frac{m \omega}{\sqrt{2mV(x')}} - \frac{1}{x'} \right)
\end{equation}
is a regularized time for a particle with $E = 0$ to cross the inverted barrier, and the $\ell_i$ are non-linear scales defined further on in \eqref{smallosclengths}. We also restrict to times $t\ll t_\textsf{dec}$, where
\begin{equation} \label{smallosctimecondition}
    \omega t_\textsf{dec}
    \equiv \frac{|\ell_3|^2}{ab e^A} \gg 1 \,.
\end{equation}
In this regime we claim the time-dependent current is well-approximated by
\begin{grayalign}
    j(x_T,t) &= \Gamma_0 \exp \left( 2 e^A \frac{a b}{d^2} \cos \varphi \right) \notag \label{joftfinalGarg} \\ &\times \exp \left\{ e^{2A} \left( \frac{a}{d} \right)^2 \left[ \left( \frac{b}{\ell_1} \right)^2 \cos 2 \varphi + \left( \left( \frac{b}{\ell_2} \right)^2 + \left( \frac{b}{\ell_3} \right)^2 \varphi \right) \sin 2 \varphi \right] - \frac{1}{2} \left( \frac{a}{d} \right)^2 + \cdots \right\} \,, \notag \\
    \varphi &= \omega (t-t_{b \to x_T}) \,,
\end{grayalign}
where, by \eqref{smalloscregime} and \eqref{smallosctimecondition}, the second line in $j$ is a small correction to the first. We included it because a sufficient way to satisfy \eqref{smalloscregime} is by sending $\hbar \to 0$ while keeping $a/d$ fixed. In this limit, Eq.~\eqref{joftfinalGarg} is the correct current to order $\hbar^0$ in the exponent. Notice that the $\mathcal{O}(a^2)$ correction destroys the exact periodicity of the current. This is because the $\varphi \sin 2 \varphi$ term receives a contribution from the anharmonic correction to $E_n \approx \hbar \omega(n+1/2)$. This effect is small for $t \ll t_\textsf{dec}$.

The total probability that escapes the well during one oscillation time is then approximately
\begin{equation}
    \Delta P = \sqrt{e^A \frac{b}{a}} \, \exp \left\{ 2 e^A \frac{a b}{d^2} + \left[ e^{2A} \left( \frac{b}{\ell_1} \right)^2 - \frac{1}{2} \right] \left( \frac{a}{d} \right)^2 \right\} e^{-2 S_0/\hbar} \,,
\end{equation}
since the effective tunneling time is $\Delta \varphi \sim \sqrt{d^2/ab e^A} \ll 1$.

%%%%%%%%%%%%%%%
\paragraph{Sketch of the computation}
%%%%%%%%%%%%%%%
We must solve \eqref{nstarrelation} for $n_*(t)$ at low energy. Near the false vacuum, expand
\begin{equation}
    V(x)=\frac12m\omega^2x^2+\lambda x^3+\mu x^4+\cdots \,.
\end{equation}
We find \cite{exercise2026}
\begin{align}
    S(E)&=S_0-\frac{\mathcal A(E)}{\omega}E
    -\frac{\mathcal B(E)}{2m\omega^3}E^2
    +\mathcal O(E^3\log E) \, , \label{SofEsmallE} \\
    \zeta(E)&=\zeta_0+\frac{\pi\hbar}{4}
    +t_{b\to x_T}E+\zeta_2E^2+\mathcal O(E^3) \, ,
\end{align}
where
\begin{align}
    S_0&=\int_0^b \sqrt{2mV} \, ,
    &\zeta_0&=\int_b^{x_T} \sqrt{-2mV} \, , \\
    \mathcal A(E)&=\log\frac{2b}{a_E}+A+\frac12 \, ,
    &a_E&=\sqrt{\frac{2E}{m\omega^2}} \, , \\
    \mathcal B(E)&=\Lambda\log\frac{2b}{a_E}+B+\Delta \, ,
    &\Lambda&=\frac{3(5\lambda^2-2m\omega^2\mu)}{2m^2\omega^4} \, .
\end{align}
Here $A$ was given in \eqref{Adefinition}, and
\begin{align}
    \Delta={}&-\frac{47\lambda^2}{8m^2\omega^4}
    +\frac{7\mu}{4m\omega^2}
    +\frac{3\lambda}{m\omega^2b}-\frac1{2b^2}
    -\frac{2(m\omega)^3}{\sqrt b\,[-2mV'(b)]^{3/2}} \, , \\
    B={}&\int_0^b\di x'\left[
    \frac{(m\omega)^3}{[2mV(x')]^{3/2}}
    -\frac1{x'^3}+\frac{3\lambda}{m\omega^2x'^2}
    -\frac{\Lambda}{x'}
    -\frac{(m\omega)^3}{[-2mV'(b)]^{3/2}(b-x')^{3/2}}
    \right] \, , \\
    \zeta_2={}&\frac{\sqrt{2m}}4\left\{
    \frac1{[-V'(b)]^{3/2}\sqrt{x_T-b}}
    -\frac12\int_b^{x_T}\di x'\left[
    \frac1{[-V(x')]^{3/2}}
    -\frac1{[-V'(b)(x'-b)]^{3/2}}
    \right]\right\} \, .
\end{align}

The resonant state condition gives
\begin{equation}
    \frac1\pi\int_{c_n}^{a_n} k_n
    =\frac{E_n}{\omega}+\frac{\Lambda E_n^2}{2m\omega^3}+\cdots
    =\hbar\left(n+\frac12\right) \, ,
\end{equation}
hence
\begin{align}
    E_n&=\hbar\omega\left(n+\frac12\right)
    -\frac{\Lambda\hbar^2}{2m}\left(n+\frac12\right)^2+\cdots \, , \label{smalloscenergylevels} \\
    \omega_n&=\omega\left[1-\Lambda d^2\left(n+\frac12\right)\right]+\cdots \, .
\end{align}
\eqref{complexWKBenergy} then gives
\begin{equation}
    \Gamma_n\simeq\Gamma_0\frac{Q^n}{n!}
    \exp\left[d^2(B+\Delta-\Lambda A)\left(n+\frac12\right)^2+\cdots\right] \, ,
\end{equation}
where
\begin{equation} \label{smalloscQGamma0}
    Q=2(b/d)^2e^{2A} \, , \qquad
    \Gamma_0=\omega\sqrt{\frac Q{2\pi}}\,e^{-2S_0/\hbar} \, .
\end{equation}
The leading expression $\Gamma_n\simeq\Gamma_0Q^n/n!$ agrees with \cite{garg2025}. We have expanded $S(E)$ one order further because the $\mathcal{O}(a^2)$ correction to the current exponent requires it.

Turning to \eqref{nstarrelation}, the low-energy expansions give
\begin{align}
    \omega_n\tau_n
    &=\frac12\log\frac{Q}{n+1/2}
    +d^2\left(n+\frac12\right)(B+\Delta-\Lambda A)+\cdots \, , \\
    \omega_n(t-t_n)
    &=\varphi-d^2\left(n+\frac12\right)
    \left(2m\omega^3\zeta_2+\Lambda\varphi\right)+\cdots \, .
\end{align}
It is convenient to introduce
\begin{align}
    z = \alpha\sqrt Q\,e^{-i\varphi} \,, \quad   \beta = \frac{d^2}{2}\left[ B+\Delta-\Lambda A
    +i\left(2m\omega^3\zeta_2+\Lambda\varphi\right)\right] \,.
\end{align}
Equation~\eqref{nstarrelation} becomes
\begin{equation}
    n_*\left(n_*+\frac12\right)
    =z^2\exp\left[4\beta\left(n_*+\frac12\right)\right] \, ,
\end{equation}
with solution
\begin{equation}
    n_*=z-\frac14+2\beta z^2+\cdots \, .
\end{equation}
This expansion requires
\begin{equation} \label{smalloscillationcondition}
    |z|\gg1 \,, \qquad |\beta z|\ll1 \,.
\end{equation}
These give \eqref{smalloscregime} and \eqref{smallosctimecondition}.

Substituting into \eqref{fppexpr}, we find
\begin{align}
    f_*={}&-\frac{S_0}{\hbar}
    +i\left(\frac{\zeta_0}{\hbar}+\frac\pi4-\frac\varphi2\right)
    +\frac14\log\frac Qz+z+\beta z^2+\cdots \, , \\
    f''_*={}&-\frac1{2n_*}-\frac1{2(n_*+1/2)}+2\beta+\cdots
    =-\frac1z+\cdots \, .
\end{align}
Putting this together in \eqref{Jexpressiongeneral}, with $T_*\simeq2\pi/\omega$ and $\cos(\arg k_*(x_T))\simeq1$ in the prefactor, we obtain \eqref{joftfinalGarg}, with
\begin{equation} \label{smallosclengths}
    \ell_1^{-2}=B+\Delta-\Lambda A \, , \qquad
    \ell_2^{-2}=2m\omega^3\zeta_2 \, , \qquad
    \ell_3^{-2}=\Lambda \, .
\end{equation}

%%%%%%%%%%%%%%%%%%%%%%%%%%%%%%%%%%%%%%%%%%%%%%%%
\subsection{More general coherent states} \label{generalcoherentsec}
%%%%%%%%%%%%%%%%%%%%%%%%%%%%%%%%%%%%%%%%%%%%%%%%
Until now we studied the decay of coherent states defined by the one-parameter family \eqref{cncoherent}. Another class of coherent states is the two-parameter family
\begin{equation} \label{cngeneralcoherent}
    c_n = \frac{1}{(2 \pi \sigma^2)^{1/4}} e^{-(n-\bar{n})^2/4\sigma^2} \,,
\end{equation}
centered at mean energy $E_{\bar{n}}$.\footnote{\label{Gaussianfootnote}Our choice \eqref{cncoherent} at large $\alpha \gg 1$ corresponds to $\bar{n} = \sigma^2 = \alpha^2$, and the Gaussian approximation is valid for $|n-\alpha^2| \ll \alpha^2$. For our saddle point \eqref{n0equation}, this implies $\omega_0 \tau_0 \ll 1$. When $\omega_0 \tau_0 = \mathcal{O}(1)$, non-Gaussian corrections must be taken into account, as we have done.} One checks that for some time $t \ll t_{\textsf{dec}} \sim \hbar/(|E''_{\bar{n}}| \sigma^2)$ this represents a localized particle in position space, with expectation value $\langle x \rangle_t$ following the classical evolution at energy $E_{\bar{n}}$ and variance
\begin{align}
    w_t^2 = \langle x^2 \rangle_t - \langle x \rangle_t^2 &= \frac{k_{\bar{n}}(\langle x \rangle_t)^2}{4 \sigma^2 m^2 \omega_{\bar{n}}^2} \left( 1 + 4 \sigma^4 \omega_{\bar{n}}^2 (t'_{c_{\bar{n}}\to \langle x \rangle_t})^2 \right) \label{wtformula} \\
    &\overset{\textsf{\tiny HO}}{=} \frac{a^2}{4 \sigma^2} \sin^2 \omega t + \frac{\sigma^2 d^4}{a^2} \cos^2 \omega t \,,
\end{align}
where in the last line we have substituted the harmonic oscillator parameters and classical motion $\langle x \rangle_t = a \cos \omega t$ for comparison.\footnote{So the profile $|\psi(x,t)|^2$ ``breathes" around $x = \langle x \rangle_t$ in general -- but it still represents a localized oscillating blob if the width $w_t$ is much smaller than the oscillation amplitude. For the harmonic oscillator, the breathing is indeed seen to be constant when $\sigma = a/(\sqrt{2} d)$, previously called $\alpha$.} So, inside the well, $\psi(x,t)$ defined by coefficients \eqref{cngeneralcoherent} is approximately a displaced Gaussian in position space with variance $2 w_t^2$ (times a phase), following the classical motion.

It is then straightforward to compute the current on the other side of the barrier, following the methods of \S\ref{decompsec} or \S\ref{PIsec} \cite{exercise2026}. Near the time $t_0$ of maximal tunneling we find
\begin{align}
    j(x_T,t) &= \frac{\omega_0}{|f''_0| \sqrt{2 \pi \sigma^2}} e^{2 \, \textsf{Re} f_0} e^{\textsf{Re}(1/f_0'') \omega_0^2 (t-t_0)^2} \,, \\
    \textsf{Re} \, f_0 &= - (\sigma \omega_0 \tau_0)^2 - \frac{S_0}{\hbar} \,, \quad f_0'' = - \frac{1}{2\sigma^2} + \omega_0(\tau'_0 + i t'_0) + \omega'_0 \tau_0 \,,
\end{align}
where $n_0$ satisfies the relation
\begin{equation}
    n_0 = \bar{n} + 2 \sigma^2 \omega_0 \tau_0 \,,
\end{equation}
and
\begin{equation}
    t_0 \equiv t_{b_0 \to x_T} \,.
\end{equation}
These expressions match \eqref{Jfinaltime}-\eqref{fpp0eq} in the Gaussian regime (see footnote~\ref{Gaussianfootnote}).

\clearpage
%%%%%%%%%%%%%%%%%%%%%%%%%%%%%%%%%%%%%%%%%%%%%%%%
\section{Numerics} \label{sec:Numerics}
%%%%%%%%%%%%%%%%%%%%%%%%%%%%%%%%%%%%%%%%%%%%%%%%
Imposing purely outgoing boundary conditions directly at a point to the right of the barrier is numerically difficult. 
Instead, we add a complex absorbing potential (CAP) in the free region far to the right of the barrier. 
This adds a small imaginary component to the potential that absorbs the outgoing wave and thereby effectively enforces outgoing boundary conditions \cite{RissMeyer1993}.
We discretize space and diagonalize this complex Hamiltonian to obtain its eigenstates and complex energies (an efficient algorithm to diagonalize complex symmetric matrices can be found in \cite{Noble:2013nr,NOBLE2017304}). Low-lying states closely resemble the eigenstates of the harmonic oscillator, while higher energy states exhibit oscillatory behavior to the right of the barrier.

\begin{figure}[h!]
    \centering
    \includegraphics[width=0.75\linewidth]{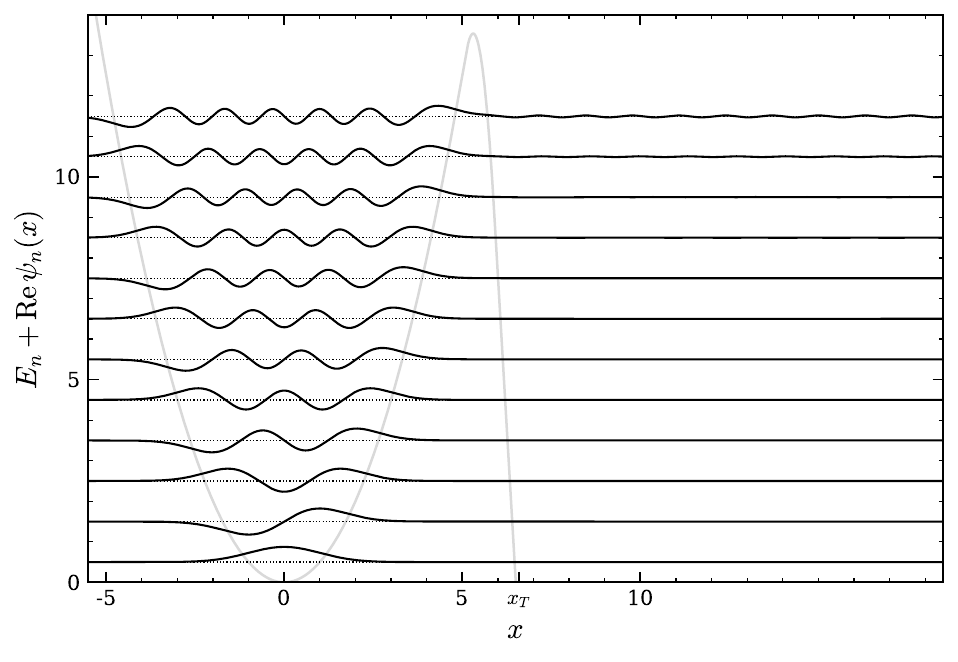}
    \caption{The first 12 resonant states as found from the CAP (Complex Absorbing Potential) Hamiltonian.}
    \label{fig:CAPstates}
\end{figure}

In our numerical study we considered the potential and parameters 
\begin{equation}
V(x) =
\begin{cases}
\frac{1}{2} m \omega^2 x^2 & x \le L \\[6pt]
V_b \left(1 - \frac{x - L}{w} \right) & L \le x \le L + w \\[6pt]
0 & L + w \le x \le x_{\mathrm{CAP}} \\[6pt]
-i \eta (x - x_{\mathrm{CAP}})^2 & x \ge x_{\mathrm{CAP}} \,,
\end{cases}
\end{equation}
with $m=\hbar = \omega = 1$, $L = 6$, $w=0.9$, 
$x_{\text{CAP}} = L + w + 50$ and 
$\eta   = 3 \times 10^{-4}$. The potential has a kink at $x=L$ where the harmonic well meets the linear ramp. We numerically smoothed this junction over a region of width $ 2 \delta $, using a cubic Hermite interpolant that matches the value and first derivative of both pieces at the endpoints of the smoothing interval. In our examples, we take $\delta = 0.3$.
The first 12 resonant states for this system are shown in Fig.~\ref{fig:CAPstates}. 

The main result we would like to verify numerically is Eq.~(\ref{formula}), in particular that it can be approximated analytically in the semiclassical limit as in \S\ref{WKBsec}-\S\ref{coherentsec}. To do this, we take an initial state which is a coherent superposition of the harmonic oscillator energy eigenstates with coefficients given in Eq.~\eqref{cncoherent}. We use the Crank-Nicolson method to evolve the initial state in time with the complex Hamiltonian.

\begin{figure}[h!]
    \centering
    \includegraphics[width=1\linewidth]{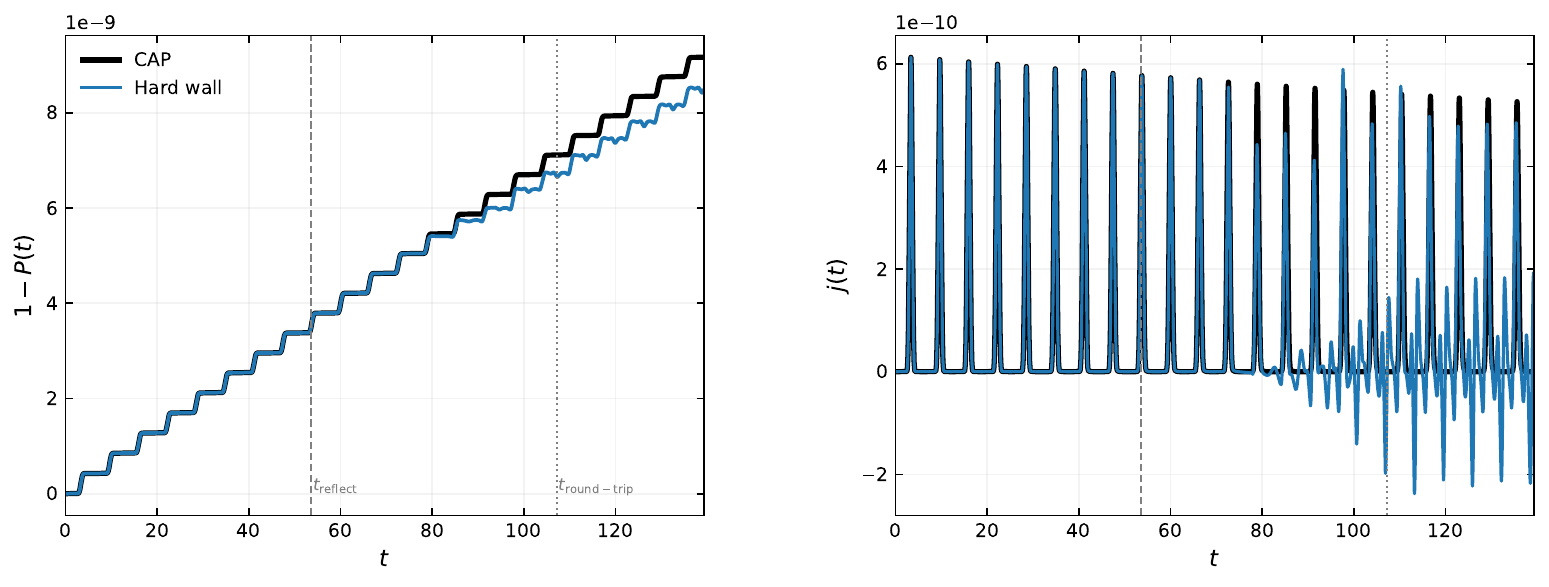}
    \caption{The tunneling probability, $1-P(t)$ (left) and probability current $j(t)$ (right) comparing the time evolution of a coherent harmonic oscillator initial state with $\alpha = 1.1$ via either CAP  (Complex Absorbing Potential) evolution (black) or evolution with hard-wall boundary conditions (blue). For early-enough times, before the outgoing wave reaches the boundary, the plots are nearly indistinguishable.}
    \label{fig:CAPvsHardwall}
\end{figure}

To validate that the CAP correctly implements outgoing boundary conditions, we compare this evolution to a reference evolution using the real Hamiltonian with hard-wall boundary conditions. At each time step we compute the probability current at $x_T$ directly from the numerically evolved wave functions and the probability to remain in the well $P(t)$ by integrating $|\psi(x,t)|^2$ over the metastable region. Close agreement is exhibited for each of these until the outgoing wave reaches the boundary and reflects back into the well, after which the hard-wall result develops small oscillations due to interference, leading to a discrepancy, as shown in Fig.~\ref{fig:CAPvsHardwall}. This verifies that we are justified in using the CAP Hamiltonian as a proxy for the true open system with an initial state that is fully localized in the well.

Now we compare the numerical CAP evolution to Eq.~\eqref{formula}. When we evaluate the formula numerically (i.e., by using the complex energies from the CAP Hamiltonian), the plots are indistinguishable by eye. This confirms both the validity of the resonant-state decomposition and the correctness of \eqref{formula}. We can also evaluate \eqref{formula} using the harmonic oscillator energies for $E_n$, the WKB approximations for $\Gamma_n$ and $\zeta_n$ given in Eqns.~\eqref{complexWKBenergy}-\eqref{AnphaseSC}, and taking the imaginary part of the momentum $k_n$ to be zero. This approximation is analytic in the sense that it does not require any knowledge of the exact eigenstates of the system or the numerical solution of the Schr\"odinger equation. 

\begin{figure}[h!]
    \centering
    \includegraphics[width=1\linewidth]{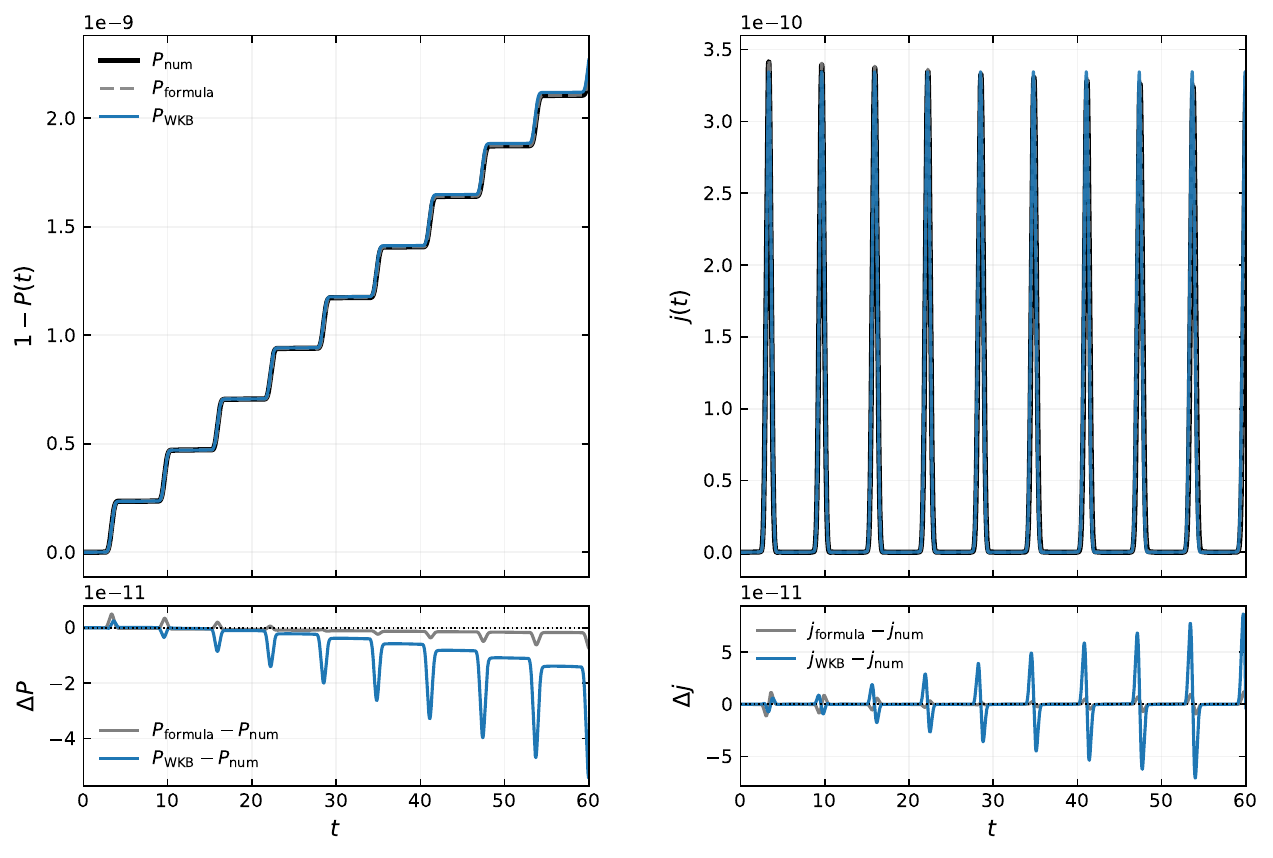}
    \caption{The tunneling probability $1-P(t)$ (top left) and probability current $j(t)$ (top right) as functions of time, for a coherent simple harmonic oscillator initial state in the potential well with $\alpha = 1.1$. The black solid lines show the direct numerical result from Crank-Nicolson time evolution with the CAP  (Complex Absorbing Potential) Hamiltonian, the gray dashed lines show the formula of Eq.~(\ref{formula}) evaluated numerically, and the blue dashed lines show the fully analytic WKB prediction using only the potential parameters. The residual panels (bottom) show the deviation of each approximation from the numerical result. The formula residual lies at the level of numerical precision, confirming the validity of the resonant-state decomposition, while the WKB residual reflects the error introduced by the analytic approximations to the decay rates and amplitudes.}
    \label{fig:WKBvsnumerics}
\end{figure}

For the time evolution of the coherent initial state, the WKB approximation agrees very well with numerics, as shown in Fig.~\ref{fig:WKBvsnumerics}.
For this numerical example we took $\alpha = 1.1$, which means that the distribution of the coefficients $c_n$ peaks at $n = 1$ and fall rapidly for larger $n$. Nevertheless, because decay is weighted by $|c_n|^2\Gamma_n$ and $\Gamma_n$ grows rapidly with $n$, even though $|c_9|^2 \approx 4.6 \times 10^{-6}$ the levels that contribute most have $n_* \approx 9$.  The tunneling action  is $S_9/\hbar\simeq5.4$ and the energy is well below the barrier peak at $E_9 = 9.5 < V_b \approx 18$.

The step structure in the tunneling probability and the corresponding spikes in the probability current have a simple physical explanation. The coherent initial state oscillates back and forth in the well, and the tunneling rate is sharply peaked when the expectation value of the position is near the right of the well, closest to the barrier. One can think of the tunneling as happening primarily at a single classical turning point, or the tunneling process being effectively localized in time near this point. 

\begin{figure}[!ht]
    \centering
    \includegraphics[width=1\linewidth]{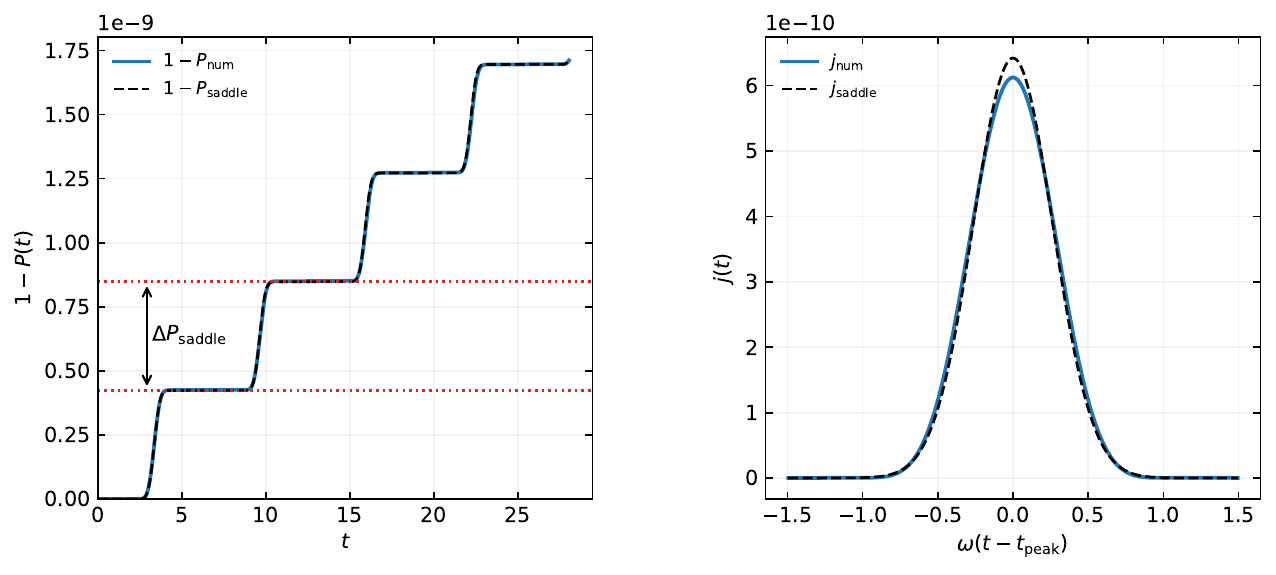}
    \caption{Numerical verification of Eqns.~\eqref{Jexpressiongeneral} and \eqref{deltaP} for a coherent initial state with $\alpha = 1.1$. Left: cumulative tunneling probability $1 - P(t)$ for the first few oscillations obtained by integrating the numerical current (blue) and the saddle-point approximation to the current from Eq.~\eqref{Jexpressiongeneral} (dashed black). The horizontal red lines mark the total $\Delta P$ per oscillation cycle predicted by Eq.~\eqref{deltaP}. Right: the probability current $j(t)$ over one oscillation period centered at the tunneling peak. The saddle-point approximation from Eq.~\eqref{Jexpressiongeneral} (dashed black) matches the numerical current; the peak and Gaussian width agree to within $4 \%$.}
    \label{fig:saddle}
\end{figure}

This structure admits a semiclassical description in which the current is dominated by a saddle-point contribution, computed in \S\ref{coherentsec}. In this approximation, the probability current near each tunneling event takes the form of a Gaussian. As shown in Fig.~\ref{fig:saddle}, the numerical current is sharply peaked in this region and is well-described by the saddle-point expression in Eq.~\eqref{Jexpressiongeneral}, with both the peak amplitude and width in good agreement. Integrating over a single oscillation period yields a total leaked probability consistent with Eq.~\eqref{deltaP}, confirming that the tunneling is concentrated in narrow peaks in time, with amplitude controlled by the semiclassical action.

\begin{figure}[!htbp]
    \centering
    \includegraphics[width=\linewidth]{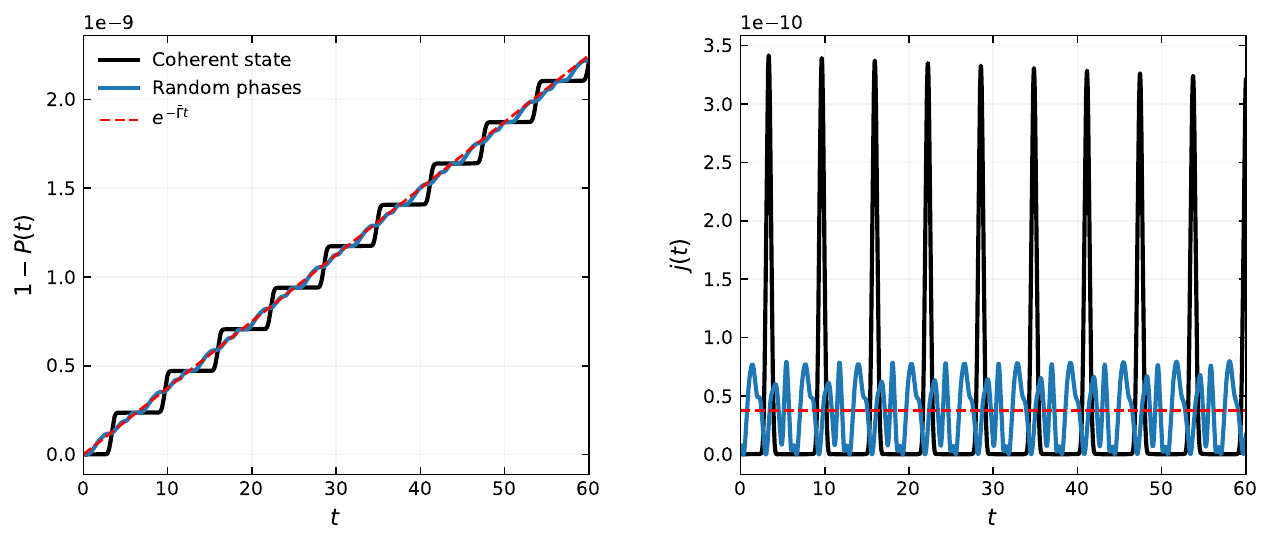}
    \caption{The cumulative tunneling probability $1-P(t)$ (left) and probability current $j(t)$ (right) for a coherent initial state with $\alpha = 1.1$ (black) and a random-phase initial state (blue), both with the same magnitude of the expansion coefficients $|c_n|$. The red dotted line shows the average decay $e^{-\bar \Gamma t} \approx 1 - \overline{\Gamma} t$ with $\bar \Gamma = \sum_n |c_n|^2 \Gamma_n$.}
    \label{fig:coherentVSrandom}

    \vspace{\floatsep}

    \includegraphics[width=\linewidth]{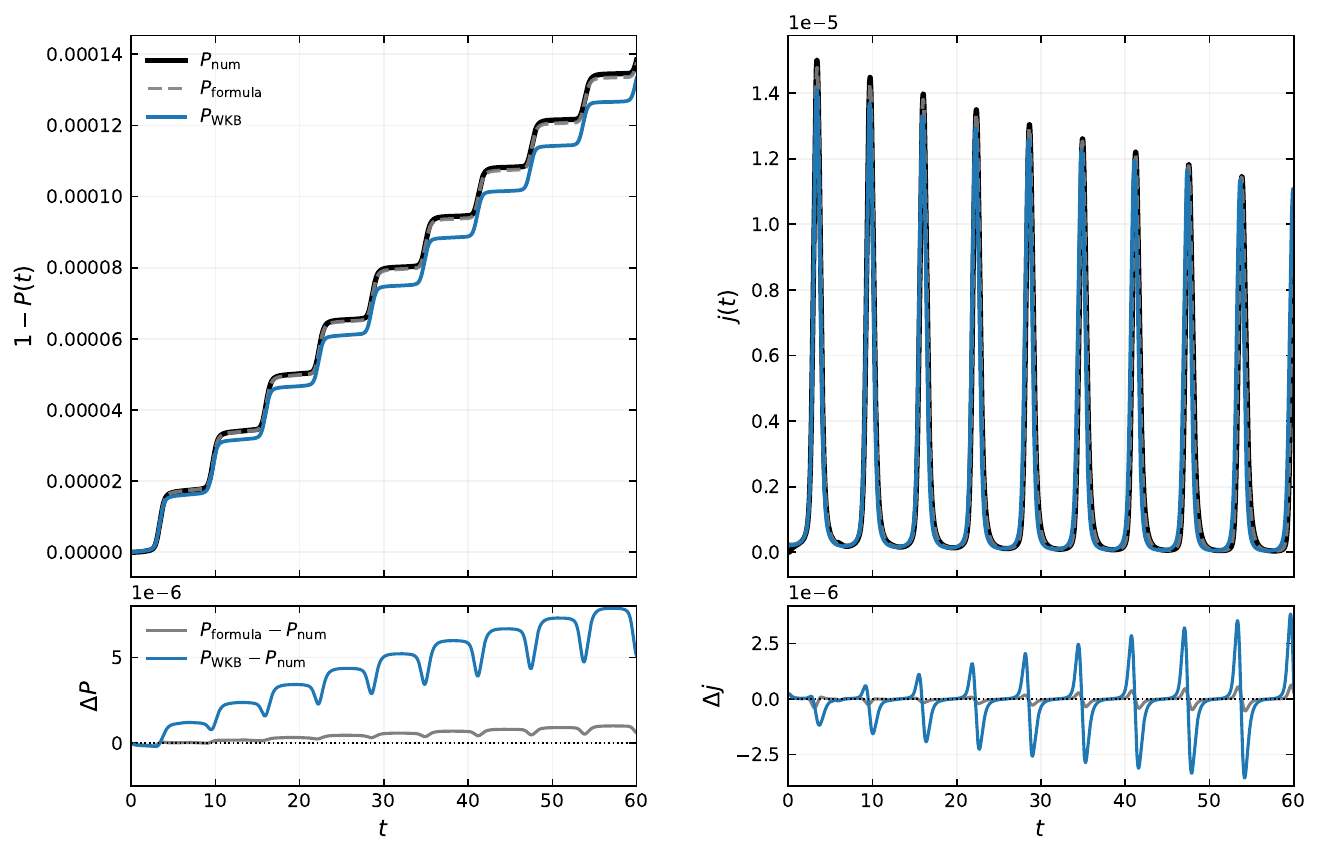}
    \caption{The tunneling probability $1 - P(t)$ (top left) and probability current $j(t)$ (top right) as functions of time, for a coherent initial state with $\alpha = 2$. Line styles are as in Fig.~\ref{fig:WKBvsnumerics}. The formula residual (bottom) remains very low, while the WKB residual is visibly larger than in Fig.~\ref{fig:WKBvsnumerics}, reflecting the reduced accuracy of the semiclassical approximation for modes with energies approaching the top of the barrier.}
    \label{fig:wkbvsnumerics_high}
\end{figure}

When the magnitudes $|c_n|$ of the expansion coefficients are held fixed but the phases $\vartheta_n$ are randomized, the coherent oscillatory motion is destroyed and the step structure disappears, leaving a decay that tracks the average rate $\bar \Gamma = \sum_n |c_n|^2 \Gamma_n $. This comparison is shown in Fig.~\ref{fig:coherentVSrandom}.

When higher energy states are considered, the resonant decomposition remains accurate, but less so than in the lower energy case. While Eq.~(\ref{formula}) remains formally valid -- and is exact for true eigenstates of $\widetilde H$ -- the practical accuracy of the decomposition is reduced because the initial state has increased overlap with higher-energy modes that are less well-described by isolated resonances.

The WKB approximation starts to noticeably deviate from the numerical answer for oscillations with larger amplitude. For $\alpha=2$, although $|c_n|^2$ peaks at $n=4$, the  WKB estimate places the peak of $|c_n|^2\Gamma_n$ near $n=n_*=14$, with $E_{14}/V_b\simeq0.8$. The smaller tunneling action of these escape-dominating modes makes WKB less reliable. The step structure in the tunneling probability is also less sharp than in the low-energy case. This is because the higher energy modes have wave functions which are noticeably distorted by the barrier. Both of these effects are shown in Fig.~\ref{fig:wkbvsnumerics_high}.

\newpage
%%%%%%%%%%%%%%%%%%%%%%%%%%%%%%%%%%%%%%%%%%%%%%%%
\section{Discussion} \label{sec:Discussion}
%%%%%%%%%%%%%%%%%%%%%%%%%%%%%%%%%%%%%%%%%%%%%%%%
A main result of this paper is Eq.~\eqref{formula}, a closed-form expression for the time-dependent probability current flowing through a potential barrier for an initial state localized in a metastable well. All ingredients in the equation (the resonant energies $E_n$, decay widths $\Gamma_n$ and phases $\zeta_n$) can be computed analytically in the semiclassical regime without knowledge of the exact energy eigenstates of the system or a numerical solution to the Schr\"odinger equation. This places our result in favorable contrast with existing approaches: path integral methods~\cite{Krais-Vakkuri_1996, Darme:2019ubo} disagree at the level of the exponent in quantum field theory, while Refs.~\cite{Lin:2025bjn,Lin:2025wgc} left several coefficients unfixed which had to be fit against a numerical solution. Our result is fully analytic and additionally includes the first subleading order in the semiclassical limit. A path integral derivation of the ingredients featuring in \eqref{formula} appeared in \cite{Janssen:2026ybl}.

A striking feature of the result is the time structure of the tunneling 
current when the initial state is a coherent superposition of energy eigenstates. Rather than decaying 
smoothly at a constant exponential rate, the current is concentrated in narrow spikes 
separated by one oscillation period $2\pi/\omega$. This structure has a simple 
semiclassical interpretation: the coherent state undergoes classical oscillations in the 
well, and tunneling is exponentially enhanced whenever the wave packet reaches the 
classical turning point nearest to the barrier. In the saddle-point approximation of \S\ref{coherentsec}, each tunneling event is well-described by a Gaussian in time with 
width $\Delta t \sim (\omega\sqrt{n_0})^{-1} \ll 2\pi/\omega$, confirming that the 
tunneling process is effectively localized in time. The total probability leaked per 
oscillation, Eq.~\eqref{deltaP}, is controlled by the semiclassical action 
evaluated at the energy of a dominant resonance. This resonance is determined by a balance between the particle having less probability to be located higher up the barrier, but having a larger probability to tunnel from there. This offers a precise quantum-mechanical realization of the intuitive notion that a particle in the well ``attempts'' to tunnel once per bounce off the barrier. A second key result of this paper is an independent path integral derivation in \S\ref{PIsec} of the time-dependent current.

The comparison between the coherent and random-phase initial states in Fig.~\ref{fig:coherentVSrandom} 
illustrates the key role of  interference. When the phases 
$\vartheta_n$ of the expansion coefficients are randomized with the magnitudes $|c_n|$ fixed, the cross-terms in Eq.~\eqref{formula} average to zero over many 
oscillation periods, and the current approaches the incoherent average 
$\overline{\Gamma} = \sum_n |c_n|^2 \Gamma_n$. The step structure and spike pattern are 
entirely due to phase coherence among the resonant states.

There are several limitations of our result. First, the resonant-state 
expansion is not complete: a full spectral decomposition requires contributions from a 
non-resonant continuum. For initial states well-localized in the well with energies well below the barrier, this contribution is negligible and the 
expansion~\eqref{resexp} is an excellent approximation. However, the formula 
breaks down at early times $t \lesssim 1/\omega$, where cancellations between 
resonant and non-resonant components are responsible for the correct $j \sim t$ 
behavior. Eq.~\eqref{formula} should therefore be understood as 
accurate only after a brief initial transient of order one oscillation period. Similarly, 
at very late times $t \gtrsim 8S_0/(\hbar\Gamma_0)$ the succession of long periods of exponential decay ceases and becomes a power-law tail.

Second, the WKB approximation to the decay widths $\Gamma_n$ becomes less accurate as 
the energy of the relevant resonances approaches the top of the barrier. This is 
illustrated in Fig.~\ref{fig:wkbvsnumerics_high}: for $\alpha = 2$, where the dominant modes have 
$E_n/V_b \sim 0.8$, the WKB residual is visibly larger than in the low-amplitude case, 
though the resonant-state formula~\eqref{formula} itself (evaluated with exact 
numerical complex energies) remains accurate throughout. The semiclassical 
approximation is most reliable when the tunneling action $S_n/\hbar \gg 1$, which 
requires the relevant energies to be well below the barrier height.

The experimental context of superconducting Josephson junctions~\cite{PhysRevLett.55.1543, PhysRevB.35.4682} offers a natural arena in which to confront these results. In those 
systems, a current-biased junction supports macroscopic quantum tunneling out of a 
tilted washboard potential, and the decay rate has been measured to depend sensitively 
on the initial quantum state. The step structure in the tunneling probability and the 
oscillatory current spikes predicted by \eqref{formula} have a direct analog in this setting: 
an initial coherent superposition of junction states would be expected to tunnel 
preferentially at definite phases of the washboard oscillation, producing a 
time-structured escape rate rather than a simple exponential. Whether this structure 
is resolvable in current experiments, which would require both phase control over the initial state and sufficient time resolution of the decay, is a question we leave for future work.

Finally, the most important extension of this work is to quantum field theory, where the 
analogous problem of tunneling from a non-vacuum initial state is both physically important and technically interesting. There are many scenarios in cosmology where this is relevant.  An oscillating axion-like field near a local minimum of its potential will decay via bubble nucleation at a rate that our results suggest is periodically enhanced, concentrated near the phases of oscillation at which the field approaches the barrier. While this calculation has been attempted independently by at least two groups \cite{Krais-Vakkuri_1996,Darme:2019ubo}, the results of this work for the analogous quantum-mechanical problem suggest that the answer remains elusive. The path integral formulation of Ref.~\cite{Janssen:2026ybl}, along the lines of \S\ref{PIsec}, may provide a useful entry point. We leave a systematic treatment of the quantum field theory case to future work.

%%%%%%%%%%%%%%%%%%%%%%%%%%%%%%%%%%%%%%%%%%%%%%%%
\section*{Acknowledgements} \label{sec:acknowledgements}
%%%%%%%%%%%%%%%%%%%%%%%%%%%%%%%%%%%%%%%%%%%%%%%%
We thank Rudin Petrossian-Byrne, Mehrdad Mirbabayi and Giovanni Villadoro for discussions.

\clearpage
%%%%%%%%%%%%%%%%%%%%%%%%%%%%%%%%%%%%%%%%%%%%%%%%
\appendix
%%%%%%%%%%%%%%%%%%%%%%%%%%%%%%%%%%%%%%%%%%%%%%%%

%%%%%%%%%%%%%%%%%%%%%%%%%%%%%%%%%%%%%%%%%%%%%%%%
\section{WKB for excited states} \label{excitedWKB}
%%%%%%%%%%%%%%%%%%%%%%%%%%%%%%%%%%%%%%%%%%%%%%%%
The Hamiltonian associated with the tunneling potential shown in Fig.~\ref{fig:potential} has a continuous, non-degenerate spectrum. In this section we focus on excited energy eigenstates with energies
\begin{equation} \label{excitedstates}
    V_0 \ll E \ll V_b \,,
\end{equation}
where the unit of energy spacing that defines the inequalities here is of order $\hbar/T(E)$, with $T(E)$ the (full) classical oscillation time at energy $E$ in the metastable region:
\begin{equation} \label{classicaloscillationtime}
    T(E) = \int_{c(E)}^{a(E)} \di x \, \frac{2m}{k(x;E)} \,, \quad k(x;E) = \sqrt{2m \left( E - V(x) \right)} \,, \quad V(c) = E = V(a) \,.
\end{equation}
So the regime \eqref{excitedstates} exists when $\hbar \ll \left( V_b - V_0 \right) \langle T \rangle$. We will abbreviate $\int f \equiv \int \di x \, f(x)$.

For energies in \eqref{excitedstates}, one can use the linear WKB turning point formulae \cite{merzbacher1998quantum} to transition between the classically allowed and forbidden regions. One starts with a purely decaying WKB branch on the left, and follows this through all the way to the free region on the right. One finds\footnote{$[1 + \mathcal{O}(\hbar)]$ correction factors to the WKB expressions are implied throughout.}
\begin{align} \label{WKBfull}
	&\psi_E(x) = \mathcal{N}_E \, \times \\
    &\left\{
    \begin{array}{ll}
        \displaystyle\frac{1}{\sqrt{\kappa}} \exp \left( - \frac{1}{\hbar} \displaystyle\int_x^c \kappa \right) & \mbox{for } x \leq c - \Delta_c \,, \\
        \displaystyle\frac{2}{\sqrt{k}} \cos \left( \frac{1}{\hbar} \displaystyle\int_c^x k - \frac{\pi}{4} \right) & \mbox{for } c + \Delta_c \leq x \leq a - \Delta_a \,, \\
            \displaystyle\frac{1}{\sqrt{\kappa}} \left[ 2 \cos \varphi \, \exp \left( \frac{1}{\hbar} \int_a^x \kappa \right) + \sin \varphi \, \exp \left( - \frac{1}{\hbar} \int_a^x \kappa \right) \right]  & \mbox{for } a + \Delta_a \leq x \leq b - \Delta_b \,, \\
        \displaystyle\frac{1}{\sqrt{k}} \left[ 4 \theta \cos \varphi \cos \left( \frac{1}{\hbar} \displaystyle\int_b^x k - \frac{\pi}{4} \right) + \frac{\sin \varphi}{\theta} \cos \left( \frac{1}{\hbar} \displaystyle\int_b^x k + \frac{\pi}{4} \right) \right] & \mbox{for } b + \Delta_b \leq x \,,
    \end{array} \notag
\right.
\end{align}
where
\begin{equation}
    \kappa(x;E) = \sqrt{2m \left( V(x) - E \right)} \,, \quad \theta(E) = e^{S(E)/\hbar} \,, \quad S(E) = \int_a^b \kappa \,, \quad \varphi(E) = \frac{1}{\hbar} \int_c^a k \,.
\end{equation}
The transition intervals $\Delta_{a,b,c}$ are chosen such that $\hbar |k'| \ll k^2 \,, \hbar |\kappa'| \ll \kappa^2$ hold at them on the one hand (so, we are sufficiently far away from the turning points), and on the other such that a linear approximation to the potential is valid at them as well (so we are not too far away from the turning points either). This is possible when
\begin{equation} \label{linearWKBconsistency}
    \hbar \ll \sqrt{m} \frac{V'(j)^2}{|V''(j)|^{3/2}} \,, \quad j \in \{ a,b,c \} \,.
\end{equation}

From \eqref{WKBfull} we see what a generic excited energy eigenstate looks like, that is, one where $\theta \cos \varphi \gg \sin \varphi / \theta$. It has a large relative support in the free region compared to the well. However, when $\varphi$ is tuned so that $\theta \cos \varphi \sim \sin \varphi / \theta$, i.e. when $\varphi \approx \pi(n+1/2)$ for integer $n$, the situation reverses and the energy eigenstates develop a large relative support in the well compared to elsewhere. We will denote the center of the resonance band by $E = E_n$, where $\varphi(E_n) = \pi (n+1/2), n \in \mathbb{N}$.

The normalization $\mathcal{N}_E$ of the energy eigenstates can be fixed by requiring
\begin{equation} \label{Enormalization}
	\langle E' | E \rangle = \int_{-\infty}^\infty \di x \, \psi_{E'}(x) \psi_E(x) = \delta(E'-E) \,.
\end{equation}
This sets\footnote{The delta function can only come from the far right regime, where the eigenfunctions are plane waves, and it is straightforward to extract this term. \textit{All} the other (finite) contributions must exactly cancel. We encourage those who have not done it to check this explicitly for a general rectangular barrier.}
\begin{equation}
	\mathcal{N}_E = \sqrt{\frac{m}{2 \pi \hbar}} \left( 4 \theta^2 \cos^2 \varphi + \frac{\sin^2 \varphi}{4\theta^2} \right)^{-1/2} \,.
\end{equation}
We are mostly interested in this expression when $E \approx E_n$. To first order in the deviation and for $\theta_n \equiv \theta(E_n) \gg 1$ we find
\begin{equation} \label{alphapert}
	\varphi(E) = \pi \left( n + \frac{1}{2} \right) + \frac{T_n}{2\hbar} (E-E_n) + \cdots \,,
\end{equation}
so
\begin{equation} \label{cossinapprox}
	\cos \varphi = -(-1)^n \frac{T_n}{2\hbar} (E-E_n) + \cdots \,, \quad \sin \varphi = (-1)^{n} + \cdots \,,
\end{equation}
and hence
\begin{equation} \label{NEapprox}
	\mathcal{N}_{E \approx E_n} = \sqrt{\frac{m \hbar}{2 \pi T_n^2}}~ \frac{1}{\theta_n}  \frac{1}{\sqrt{(E-E_n)^2 + \frac{\hbar^2}{4 T_n^2 \theta_n^4}}} \,,
\end{equation}
where $T_n = T(E_n)$. We identify the famous Breit-Wigner distribution \cite{PhysRev.49.519}. The approximation \eqref{NEapprox} remains valid as long as $|E-E_n| \ll \hbar/T_n$, that is, for energy differences much smaller than the resonant energy level spacing. So when $|E-E_n| \lesssim \hbar/(T_n \theta_n^2)$, $\mathcal{N}_E \propto \theta_n$ is exponentially large. Outside of this regime, $\mathcal{N}_E \propto 1/\theta(E)$ is exponentially small.

%%%%%%%%%%%%%%%%%%%%%%%%%%%%%%%%%%%%%%%%%%%%%%%%
\section{Time evolution of resonant states} \label{resonanttimeevolutionsec}
%%%%%%%%%%%%%%%%%%%%%%%%%%%%%%%%%%%%%%%%%%%%%%%%
In the main text we defined resonant states as those states at quantized complex energies $\varepsilon_n$ which decay to one side and have purely outgoing boundary conditions on the other, Eq.~\eqref{resonantenergies}. These states are not $L^2$-normalizable because they exhibit an exponential growth at large $x$. We argued, though, that we could cut them off at $x = x_T$, rendering them effectively normalizable, Eq.~\eqref{psi_norm}, and that for $x \lesssim x_T$ we could expand -- to good approximation -- an initial state localized in the well in terms of these truncated resonant states and proceed from there. The expansion works well in the semiclassical limit, but the complex resonant states above are well-defined in general.

In the semiclassical limit, another way of defining resonant states, leading to the same result, is as $L^2$-normalized versions of energy eigenstates with (real) energy $E = E_n$ satisfying the WKB quantization condition $\varphi(E_n) = \pi(n+1/2), n \in \mathbb{N}$, of the well. For highly excited states with $n \gg 1$ they are given by
\begin{align} \label{resonantfull}
	\psi_n(x) = \mathcal{N}_n \left\{
    \begin{array}{ll}
        \displaystyle\frac{1}{\sqrt{\kappa_n}} \exp \left( - \frac{1}{\hbar} \displaystyle\int_x^{c_n} \kappa_n \right) & \mbox{for } x \leq c_n - \Delta_c \,, \\
        \displaystyle\frac{2}{\sqrt{k_n}} \cos \left( \frac{1}{\hbar} \displaystyle\int_{c_n}^x k_n - \frac{\pi}{4} \right) & \mbox{for } c_n + \Delta_c \leq x \leq a_n - \Delta_a \,, \\
            \displaystyle\frac{(-1)^n}{\sqrt{\kappa_n}} \exp \left( - \frac{1}{\hbar} \int_{a_n}^x \kappa_n \right)  & \mbox{for } a_n + \Delta_a \leq x \leq x_* \,, \\
        \text{decays further} & \mbox{for } x \geq x_* \,.
    \end{array}
\right.
\end{align}
Here $x_*$ is a point in the barrier where the wave function has become exponentially small. The normalization $\mathcal{N}_n$ can be calculated at large $n$ by noting that the main contribution to $|| \psi_n ||_2$ comes from the classically allowed region. The rapid oscillations of $\cos$ in this region eliminate half of the integral, and we find
\begin{equation}
    \mathcal{N}_n = \sqrt{\frac{m}{T_n}} \,.
\end{equation}

To study the time evolution of this state, $\psi_n(x,t)$ where $\psi_n(x,t=0) = \psi_n(x)$, we decompose it into energy eigenstates. Given our normalization \eqref{Enormalization}, we have
\begin{align} \label{psixt}
	\psi_n(x,t) = \int_0^\infty \di E \left( \int_{-\infty}^{\infty} \di x' \, \psi_E(x') \psi_n(x') \right) \psi_E(x) \, e^{-iEt/\hbar} \,.
\end{align}
The coefficient of $\psi_E(x)$ in the integrand is negligible unless $E \approx E_n$, and the main contribution to the integral comes from the classically allowed region. There we have $\psi_E(x') = \left( \mathcal{N}_{E \approx E_n}/\mathcal{N}_n \right) \psi_n(x')$ and so
\begin{align} \label{psixt2}
	\psi_n(x,t) \approx \int_0^\infty \di E \, \frac{\mathcal{N}_{E \approx E_n}}{\mathcal{N}_n} \psi_E(x) \, e^{-iEt/\hbar} \,.
\end{align}
We will focus on locations $x > b_n$ in the free region. Using the expression for the excited energy eigenstates of Appendix~\ref{excitedWKB} for $E \approx E_n$, together with \eqref{cossinapprox}, we find
\begin{equation} \label{psiXToutsideHighEnergy}
	\psi_n(x,t) = \sqrt{\frac{m}{T_n}} \frac{(-1)^n}{\sqrt{k_n(x)}} \exp \left( \frac{i}{\hbar} \int_{b_n}^x k_n + \frac{i \pi}{4} \right) e^{-S_n/\hbar} e^{-i E_n t/\hbar} e^{-\Gamma_n t/2} \,, \quad \Gamma_n = \frac{1}{T_n} e^{-2 S_n/\hbar} \,.
\end{equation}

%%%%%%%%%%%%%%%%%%%%%%%%%%%%%%%%%%%%%%%%%%%%%%%%
\section{WKB for low-lying states} \label{lowlyingWKB}
%%%%%%%%%%%%%%%%%%%%%%%%%%%%%%%%%%%%%%%%%%%%%%%%
Low-lying energy eigenstates are those with $E \gtrsim V_0$ for which \eqref{linearWKBconsistency} does not hold, and so the linear connection formulae cannot be applied. Approximating the potential as quadratic near the bottom,
\begin{equation} \label{quadraticwell}
    V(x) = \frac{1}{2} m \omega^2 x^2 + \mathcal{O}(\lambda x^3) \quad \text{as } x \to 0 \,,
\end{equation}
so that the classical turning points are at $a = -c = \sqrt{2E/m\omega^2}$, low-lying states have $E = \mathcal{O}(\hbar \omega)$.\footnote{Therefore our neglect of the correction to \eqref{quadraticwell} is valid when $\hbar \ll m^3 \omega^5/\lambda^2$.} In this case one can use quadratic turning point formulae \cite{Berry:1972na}. We briefly review the computation (see also \cite{garg2025}): an energy eigenstate with energy $E$ in the potential \eqref{quadraticwell} satisfies
\begin{equation} \label{eSchr}
	-\partial_y^2 \psi_E + y^2 \psi_E = (1+2n) \psi_E \,,
\end{equation}
where
\begin{equation} \label{energydefs}
	E = \hbar \omega \left( n + 1/2 \right) \,, \quad n \in \mathbb{R} \,, \quad y \equiv \frac{x}{d} \,, \quad d \equiv \sqrt{\frac{\hbar}{m \omega}} \,.
\end{equation}
We stress that $n$ here is a \textit{real number}, because there is a continuum and we are considering a general energy $E = \mathcal{O}(\hbar \omega)$. Two linearly independent solutions to \eqref{eSchr} are
\begin{equation}
	\chi_n = e^{-y^2/2} H_n(-y) \,, \quad \xi_n = e^{y^2/2} H_{-n-1}(iy) \,,
\end{equation}
where $H_n$ are Hermite ``polynomials" (which are only actual polynomials for integer $n$, and more generally have an expression in terms of $_1 F_1$), so the general solution is
\begin{equation}
	\psi_E = c_1 \chi_n + c_2 \xi_n \,.
\end{equation}
$\chi_n$ is real for real $y$ -- also for non-integer $n$ -- and has asymptotic expansions
\begin{align}
	&\chi_n(y) = e^{-y^2/2} (-2y)^n \left[ 1 + \mathcal{O}(y^{-2}) \right] \quad \text{as } y \to -\infty \,, \label{quadatyinf} \\
	&= -e^{y^2/2} y^{-n-1} \frac{\Gamma(1+n) \sin(\pi n)}{\sqrt{\pi}} \left[ 1 + \mathcal{O}(y^{-2}) \right] + e^{-y^2/2} (2y)^n \cos(\pi n) \left[ 1 + \mathcal{O}(y^{-2}) \right] \quad \text{as } y \to \infty \,, \notag
\end{align}
so that, simply, the energy eigenstates in our problem, where the potential increases indefinitely to the left, have $c_2 = 0$. Notice that if $n \in \mathbb{N}$ we recover the usual quantized harmonic oscillator energy eigenstates, which also fall off as $y \to +\infty$.

Now, far inside the barrier but still within the quadratic approximation, $d \ll x \ll m \omega^2/\lambda$, the usual WKB approximation becomes valid:
\begin{equation}
	\hbar \frac{|\kappa'|}{\kappa^2} = \left( \frac{d}{x} \right)^2 \ll 1 \,.
\end{equation}
In this region we can express the energy eigenstate as
\begin{align}
	\psi_E &= \frac{\mathcal{N}'_E}{\sqrt{\kappa}} \left[ A \exp \left( \frac{1}{\hbar} \int_a^x \kappa \right) + \exp \left( - \frac{1}{\hbar} \int_a^x \kappa \right) \right] \,, \label{psiEWKB}
\end{align}
for undetermined coefficients $\mathcal{N}_E',A$. Here
\begin{equation}
	a = d\sqrt{1+2n}
\end{equation}
is the location on the right inside the well where $E = V(a)$. We find
\begin{align}
	\frac{1}{\hbar} \int_a^x \kappa = \frac{y^2}{2} - \frac{1}{4}(1+2n) \left[ 1 + \log \left( \frac{4 y^2}{1+2n} \right) \right] + \mathcal{O}(y^{-2}) \quad \text{as } y \to \infty \,,
\end{align}
while $\kappa \sim \hbar y/d$ so that, after some manipulation,
\begin{align}
	&\psi_E \sim \mathcal{N}'_E \, \times \\ &\frac{\sqrt{d/\hbar}}{2^n\cos(\pi n)} \left( \frac{4e}{1+2n} \right)^{n/2+1/4} \left[ A \left( \frac{e}{n+1/2} \right)^{-n-1/2} \frac{\cos(\pi n)}{\sqrt{2}} \,  e^{y^2/2} y^{-n-1} + e^{-y^2/2} (2y)^n \cos(\pi n) \right] \,. \notag
\end{align}
Comparing with \eqref{quadatyinf} we deduce
\begin{equation} \label{ABrelation}
	A = - \sqrt{\frac{2}{\pi}} \, \Gamma(1+n) \tan(\pi n) \left( \frac{e}{n+1/2} \right)^{n+1/2} \,.
\end{equation}
Notice that
\begin{equation}
	A \sim -2 \tan(\pi n) \quad \text{as } n \to \infty \,,
\end{equation}
while, according to \eqref{WKBfull}, to which it should match in the $n \to \infty$ limit,
\begin{equation}
	A = 2 \frac{\cos \varphi}{\sin \varphi} \overset{!}{=} -2 \tan(\pi n) \,,
\end{equation}
where we used
\begin{equation}
	\varphi(n) = \frac{1}{\hbar} \int_c^a k = \pi \left( n + 1/2 \right) \,, \quad n \in \mathbb{R} \,.
\end{equation}

Starting from the form \eqref{psiEWKB} inside the barrier, in the free region to the right we find (now using the linear WKB connection formula)
\begin{align} \label{psifreegeneral}
	\psi_{E,\textsf{free}} = \frac{\mathcal{N}'_E}{\sqrt{k}} \left[ 2 A \theta \cos \left( \frac{1}{\hbar} \int_b^x k - \frac{\pi}{4} \right) + \frac{1}{\theta} \cos \left( \frac{1}{\hbar} \int_b^x k + \frac{\pi}{4} \right) \right] \,.
\end{align}
Just as before we can delta-function normalize the energy eigenstates by \eqref{Enormalization}, which sets
\begin{equation}
	\mathcal{N}'_E = \sqrt{\frac{m}{2 \pi \hbar}} \left( A^2 \theta^2 + \frac{1}{4 \theta^2} \right)^{-1/2} \,,
\end{equation}
with $A$ for $n \in \mathbb{R}$ given in \eqref{ABrelation}. As before, the normalization is exponentially enhanced when $n$ is close to an integer. Denoting momentarily the real, continuous parameter $n$ that is related to the continuous energy $E$ via \eqref{energydefs} by $n_E$, expanding near an integer $n$ gives
\begin{equation}
	A = - \sqrt{2\pi} \left( \frac{e}{n+1/2} \right)^{n+1/2} n! \left( n_E - n \right) + \mathcal{O} \left[ \left( n_E - n \right)^2 \right] \,,
\end{equation}
or
\begin{equation} \label{Aclosetoresonance}
	A = - \frac{ g_n T_n}{\hbar} (E-E_n) + \mathcal{O} \left[ (E-E_n)^2 \right] \,,
\end{equation}
where
\begin{equation}
	T_n \equiv \frac{2 \pi}{\omega} \,, \quad E_n \equiv \hbar \omega \left( n + 1/2 \right) \,, \quad g_n \equiv \frac{1}{\sqrt{2\pi}} \left( \frac{e}{n+1/2} \right)^{n+1/2} n! \,.
\end{equation}
We have $g_0 = 0.93, g_1 = 0.97, g_2 = 0.98$ with $g_n = 1 - (1/24n) + \mathcal{O}(n^{-2})$ as $n \to \infty$. With this,
\begin{align}
	\mathcal{N}'_{E \approx E_n} = \sqrt{\frac{m \hbar}{2 \pi (g_n T_n)^2}}~ \frac{1}{\theta_n}  \frac{1}{\sqrt{(E-E_n)^2 + \frac{\hbar^2}{4 (g_n T_n)^2 \theta_n^4}}} \,,
\end{align}
so that what changes compared to our previous analysis, \eqref{NEapprox}, is just $T_n \to g_n T_n$.

To summarize, we have the discrete, normalized, low-lying resonant energy eigenstates of the well,
\begin{equation}
	\psi_n(x) = \mathcal{N}_n \, e^{-(x/d)^2/2} H_n(-x/d) \,, \quad \mathcal{N}_n = \frac{1}{\sqrt{d \sqrt{\pi} \, 2^n n!}} \,, \quad n \in \mathbb{N} \,,
\end{equation}
and we have the continuous, delta-function normalized, low-lying true energy eigenstates, with expression in the quadratic region given by
\begin{equation}
	\psi_E = \mathcal{N}_E \, \chi_n(y) \,, \quad \mathcal{N}_{E \approx E_n} = \mathcal{N}'_{E \approx E_n} \frac{\sqrt{d/\hbar}}{2^n \cos(\pi n)} \left( \frac{4e}{1+2n} \right)^{n/2+1/4} \,.
\end{equation}

Finally, repeating the analysis of Appendix~\ref{resonanttimeevolutionsec} for the time evolution of resonant states yields a generalization of \eqref{psiXToutsideHighEnergy} that is accurate for all the resonant energy eigenstates of the well with energy well below the top of the barrier:
\begin{equation} \label{psiXToutsideGeneral}
	\psi_n(x,t) = \sqrt{\frac{m}{g_n T_n}} \frac{(-1)^n}{\sqrt{k_n(x)}} \exp \left( \frac{i}{\hbar} \int_{b_n}^x k_n + \frac{i \pi}{4} \right) e^{-S_n/\hbar} e^{-i E_n t/\hbar} e^{-\Gamma_n t/2} \,, \quad \Gamma_n = \frac{1}{g_n T_n} e^{-2 S_n/\hbar} \,,
\end{equation}
for $x > b_n$ in the free region.

%%%%%%%%%%%%%%%%%%%%%%%%%%%%%%%%%%%%%%%%%%%%%%%%
\bibliographystyle{klebphys2}
\bibliography{references.bib}
%%%%%%%%%%%%%%%%%%%%%%%%%%%%%%%%%%%%%%%%%%%%%%%%

\end{document}